\documentclass[12pt]{iopart}
\usepackage{graphicx}
\usepackage{booktabs}
\usepackage{iopams}
\usepackage[pdfusetitle,
bookmarks=true,colorlinks=true, linkcolor=blue, citecolor=blue, linktoc=page]{hyperref}
\usepackage{comment}
\usepackage{mhchem}
\usepackage[dvipsnames]{xcolor}
\usepackage{subcaption}
\usepackage[shortlabels]{enumitem}
\usepackage{multirow}
\usepackage{multicol}
\usepackage{amsmath}

\begin{document}
\title[]{Numerical simulation of a helium Plasma-Material Interaction experiment in GyM linear device through SOLPS-ITER and ERO2.0 codes}

\author{F. Mombelli$^{1}$, G. Alberti$^{1}$, E. Tonello$^{2}$, C. Tuccari$^{1}$, A. Uccello$^{3}$, C. Baumann$^{4}$, X. Bonnin$^{5}$, J. Romazanov$^{4}$, M. Passoni$^{1,3}$ and the GyM team$^{6}$}

\address{$^{1}$ Politecnico di Milano, Department of Energy, Milan, 20133, Italy}
\address{$^{2}$ Ecole Polytechnique Fédérale de Lausanne, Swiss Plasma Center, Lausanne, 1015, Switzerland}
\address{$^{3}$ Istituto per la Scienza e Tecnologia dei Plasmi, Consiglio Nazionale delle Ricerche, Milan, Italy}
\address{$^{4}$ Forschungszentrum Jülich GmbH, Institut für Energie- und Klimaforschung – Plasmaphysik, Partner of the Trilateral Euregio Cluster (TEC), 52425 Jülich, Germany}
\address{$^{5}$ ITER Organization, 13067 St Paul Lez Durance Cedex, France}
\address{$^{6}$ See the author list of A. Uccello et al 2023 \textit{Front. Phys.} 11:1108175.}

\ead{fabio.mombelli@polimi.it}

\providecommand{\keywords}[1]
{
  \small	
  \textbf{\textit{Keywords---}} #1
}

\begin{abstract}
Learning how to safely handle Plasma-Material Interaction (PMI) is a key challenge towards the commercialisation of energy from nuclear fusion. In this respect, linear plasma devices are ideal experimental testbeds, and numerical codes play a crucial complementary role. In this paper, a numerical investigation of PMI-relevant helium plasma experimental discharges in GyM linear device is presented, in which SOLPS-ITER and ERO2.0 codes are coupled for plasma background generation and material erosion investigation respectively, with the aim to support the interpretation and complement the available experimental dataset. On the plasma side, simulated profiles are validated against experimental data to provide a realistic plasma background, and the role of He metastable states is assessed for the first time in SOLPS simulations. On the material side, the erosion and deposition effects due to the introduction of the sample-holder in the simulation volume are investigated, now considering also the real stainless steel composition as wall material.
\end{abstract}


\section{Introduction}

The path towards the profitable implementation of a magnetic confinement nuclear fusion reactor, which sees ITER as a crucial milestone \cite{sips_assessment_2018}, raises the compelling need to address the topic of Plasma-Material Interaction (PMI) \cite{brezinsek_plasmawall_2017}. On one hand, the build-up of plasma particle and heat fluxes towards solid plasma-facing components (PFC) results in their erosion and surface modification, threatening their properties and integrity \cite{wesson_tokamaks_2011}. On the other, the transport of eroded impurities into the plasma may result in excessive fuel dilution and enhanced radiation losses, while their deposition may contribute to tritium retention and promote dust formation \cite{brezinsek_plasmawall_2017}. \par

Coherently, the investigation of PMI has gained high priority within the European fusion research programme coordinated by EUROFusion Consortium \cite{brezinsek_plasmawall_2017}: the Work Package Plasma-Wall Interaction and Exhaust (WP PWIE) is meant to pursue this objective exploiting both \textit{ad-hoc} experimental and numerical methods. \par

Although several PMI-relevant experimental campaigns have been carried out in tokamaks \cite{tsitrone_investigation_2022,hakola_gross_2021}, particle and heat fluences to PFCs in present-day devices are orders of magnitude lower than those expected in future reactors due to generally low pulse duration \cite{international2023iaea}. For steady-state ITER operation, ion fluxes $\leq 10^{25}$ ions$\cdot$ m$^{-2}$s$^{-1}$ and heat fluxes of 10 MW$\cdot$ m$^{-2}$ are expected onto the divertor tiles \cite{Merola2014}, to be integrated over the long duration of the plasma discharges ($\sim$ 1000 s) \cite{Shimomura1999}. Linear plasma devices (LPDs), thanks to their simple, cost-effective and flexible design, have thus been widely exploited to fill such experimental gaps, appearing as ideal testbeds for PMI phenomena \cite{Kreter2011, linsmeier_material_2017}. \par 

\begin{figure} 
  \centering
  \includegraphics[width=0.8\textwidth]{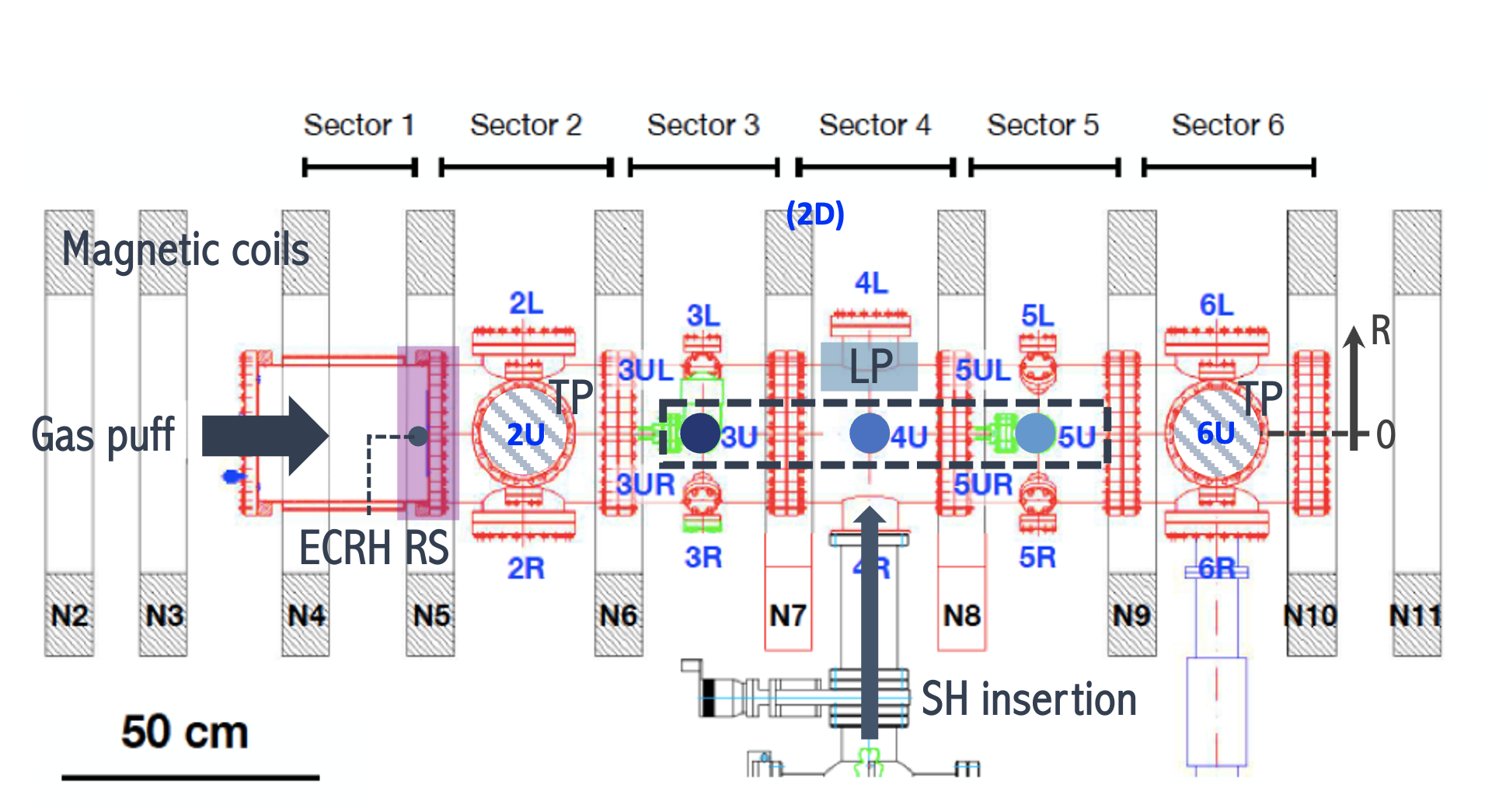}
  \caption{2D schematic of GyM LPD: the position of magnetic coils, gas puff, turbo-molecular pumps (TP, in the plane), sample-holder (SH) insertion and Langmuir probes (LP, within blue dashed box) are visible. The approximate position of the ECRH resonant surface (ECRH RS), where the external power is favourably absorbed by electrons, is pointed out on the same side of the gas puff valve. The reference radial axis is shown on the right, the origin lying on the axis of the cylinder.}
  \label{fig:2D}
\end{figure}

Among the several LPDs which operate worldwide, the present paper considers GyM linear device, hosted at Istituto per la Scienza e Tecnologia dei Plasmi - Consiglio Nazionale delle Ricerche (ISTP - CNR), Milan, whose schematic is visible in figure \ref{fig:2D} and whose characteristics are comprehensively detailed in reference \cite{uccello_linear_2023}. GyM LPD generates both hydrogenic and non-hydrogenic plasma \cite{uccello_linear_2023}: namely, the relevance of helium (He) plasma to fusion research arises from the fact that He ashes will always be present in the thermonuclear plasma as reaction products, assessing its role in PMI is therefore crucial. \par

Besides the experimental investigation, numerical codes offer the opportunity to interpret the experimental outcomes, gaining insight into the underlying physical mechanisms. Once validated, codes also acquire the potential to predict relevant scenarios which are not experimentally accessible. Moreover, the flexibility of LPD configures them as ideal testbeds for the development and validation of numerical tools \cite{uccello_linear_2023, alberti_global_2023}. It is in the groove of the numerical analysis that this contribution places itself. \par

Two state-of-the-art codes are exploited for the present modelling activity, i.e. SOLPS-ITER \cite{wiesen_new_2015, BONNIN2016} and ERO2.0 \cite{Romazanov2017}. The first one couples a 2D multi-fluid edge plasma solver (B2.5) \cite{Rozhansky2001, Rozhansky2009} with a 3D kinetic Monte-Carlo code simulating the transport of neutral species (EIRENE) \cite{reiter_eirene_2005}. Conversely, ERO2.0 is a 3D Monte-Carlo code simulating the material erosion by a plasma, the transport and deposition of the released impurities, both at a global and at a local microscopic scale \cite{Romazanov2017}. \par

Although the SOLPS package is optimised for the simulation of the Scrape-Off Layer of toroidal devices, it has already been applied to several LPDs \cite{Baeva2007,Rapp2015,Kafle2018, Zhang2022}. Concerning the specific case of non-hydrogenic plasma in GyM, SOLPS was successfully applied for both argon (Ar) \cite{sala_simulations_2020} and He plasmas \cite{tonello_point_2021}. ERO2.0 has also been applied to the study of PMI in LPDs \cite{Eksaeva2020}, while a morphology analysis at the microscale of samples exposed to a He plasma in GyM was presented in \cite{alberti_ero20_2021}. \par

More recently, a novel comprehensive numerical analysis of PMI-relevant He plasma discharges was presented, in which a coupling procedure between SOLPS-ITER and ERO2.0 codes was implemented \cite{alberti_global_2023}. In such work, the He plasma was firstly generated with SOLPS, yielding a 2D distribution of electron and ion quantities. Such distributions were eventually exploited as inputs, or background, for the ERO2.0 simulation of the global erosion of GyM internal walls, impurity transport and deposition as a function of the wall material and bias voltage applied, setting a one-way coupling between the two codes \cite{alberti_global_2023}. This work represented a first step towards the integrated plasma-material modelling of a realistic PWI experiment in GyM, yet it left several points open for further studies. Indeed, neither the plasma profiles nor the erosion outcomes were validated against experimental data. Also, neither the plasma nor the material simulations accounted for the presence of a central sample-holder and its manipulator within the geometry of the device. Finally, the erosion simulations considered the walls of the device to be made of either iron (Fe) or copper (Cu), rather than of realistic stainless steel (SS), since the ERO2.0 database did not include the sputtering yield for a He plasma on SS steel at that time. \par

The present work intends to extend the analysis initiated in \cite{alberti_global_2023}, i.e. to present a SOLPS-ITER and ERO2.0 coupled numerical investigation of plasma and material global erosion in GyM, yet partially overcoming the limitations associated to the previous study. To this purpose, realistic PMI experimental scenarios featuring He plasma discharges have been considered as a reference and exploited for the validation of SOLPS numerical results on the plasma side. For the first time, the effect of including long-lived metastable states among the neutral population of a pure He plasma in SOLPS runs is assessed. Hence, an optimised He plasma background has been used as an input for ERO2.0 simulations. In the step of defining GyM geometry for ERO2.0, a sample-holder in which samples may be arranged for exposure to plasma has been included. Erosion, transport and deposition of impurities originating both from the walls of the device and from the sample-holder have been evaluated as a function of the bias voltage applied and the wall material, now including the SS composition. \par

This work is articulated in four main sections as follows: in section \ref{reference_experimental_scenario}, the reference experimental framework on which the paper is based is described. In section \ref{sec:plasma_simulation}, the setup and major results of the SOLPS-ITER numerical investigation of the He plasma are discussed. Similarly does section \ref{sec:erosion} as concerns the ERO2.0 material erosion simulation. In section \ref{sec:conclusion}, the main outcomes of the work are recapped, and possible continuations are discussed.

\section{Reference experimental framework}
\label{reference_experimental_scenario}
A series of six PMI-relevant GyM discharges are considered as a reference experimental framework for the validation of plasma simulations. They all feature He as the main plasma species, with a He puffing strength of 42 sccm provided through the inlet valve at one of the bases of the SS cylindrical vessel (radius and length of $R_0=0.125$ m and $L_0=2.11$ m). Two turbo-molecular pumps guarantee a pumping speed of 500 L/s each, and external power is supplied by one of the two available ECRH gyrotron sources at 2.45 GHz and 3 kW nominal power ("SM" source with reference to \cite{uccello_linear_2023}, connected to port 2U in figure \ref{fig:2D}).

The axial magnetic field is generated by the surrounding coils, and the six experiments are categorised and labelled into three couples according to the intensity of the current flowing in such coils ($I_{coil}$), of 560 A, 600 A and 640 A respectively. Experimentally, the coil current configuration of 600 A is the one usually exploited for conducting PMI experiments in GyM \cite{Sala2020, Uccello2020, Uccello2023}. Conversely, the values of 560 A and 640 A coil current are the minimum and the maximum ones allowing to have the resonant surfaces located inside the vacuum vessel: experiments at these $I_{coil}$ were conducted for plasma characterisation through spectroscopic techniques and Langmuir probes rather than for PMI.

For each value of $I_{coil}$, two experiments were executed, in the presence and absence of a sample-holder (SH) in which samples for PMI studies could be arranged (see figure \ref{fig:SHphoto} below). The electron density $n_e$ and temperature $T_e$ radial profiles were acquired by three Langmuir probes (LP) located at axial positions 3U, 4U and 5U respectively (see figure \ref{fig:2D}). When present, the SH was inserted on the axis of the cylinder by means of a manipulator in correspondence of axial position 4U, thus replacing the corresponding LP.

\begin{figure} 
  \centering
  \includegraphics[width=\textwidth]{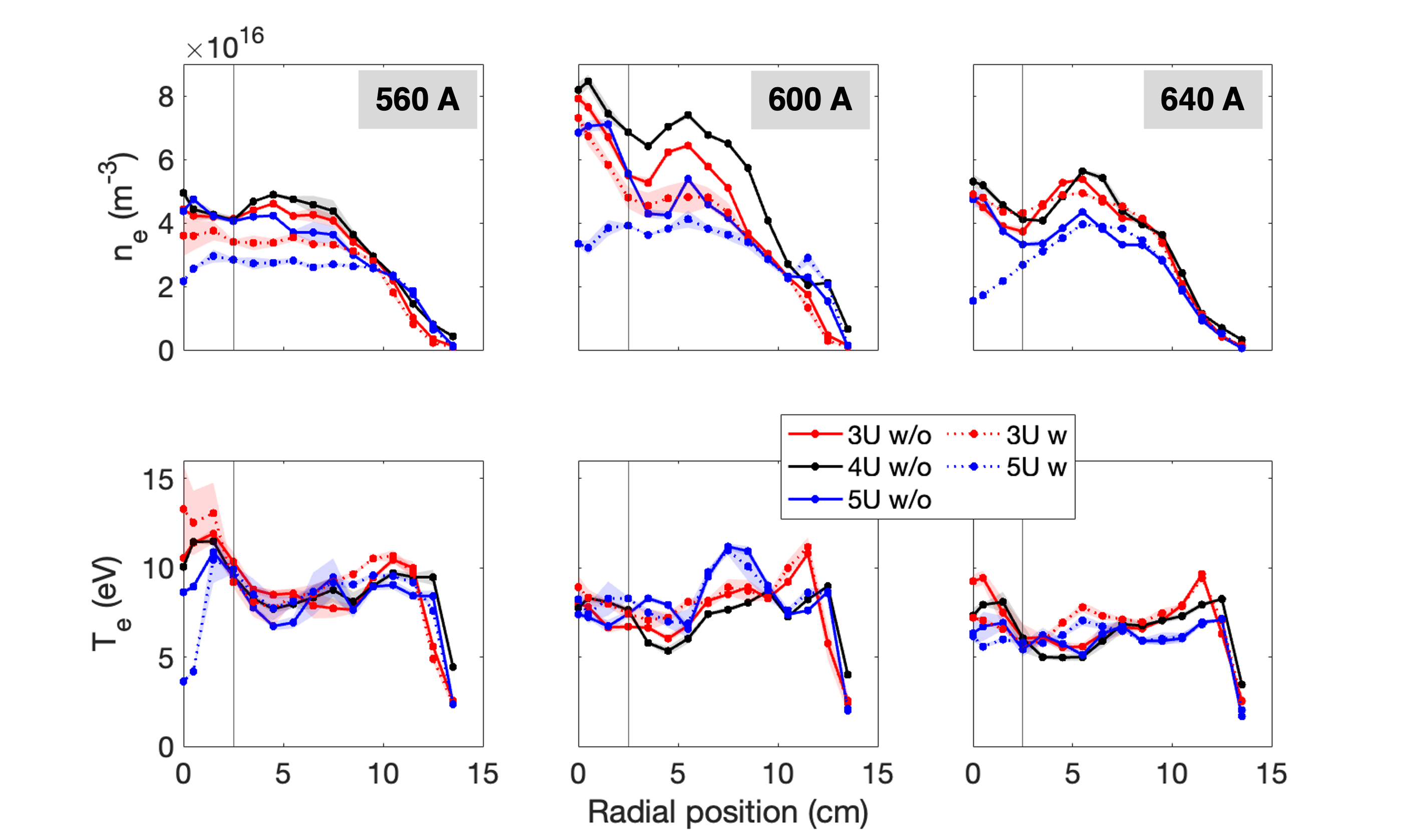}
  \caption{Experimental profiles of electron density ($n_e$, first row) and temperature ($T_e$, second row) acquired by Langmuir probes (LP), as a function of the radial coordinate (spanning from the cylinder axis to the lateral wall as depicted in figure \ref{fig:2D}). Each column is associated with one of the three values of $I_{coil}$. Each colour is associated with the axial position of one of the three LP (see figure \ref{fig:2D}), while the line style refers to the presence (dotted) or absence (solid) of the SH respectively. The thin grey vertical line represents the boundary of the SH (where present), which extends from the axis of the device (R = 0) to R $\sim$ 2.5 cm.}
  \label{fig:exp_data}
\end{figure}

At present, the only available experimental data concern plasma parameters: no measurements of the global erosion and deposition within the GyM vessel are available at this stage. Figure \ref{fig:exp_data} shows the experimental $n_e$ and $T_e$ radial profiles for the three couples of discharges at the positions of LPs. In all cases, the electron density is in the order of $\sim 10^{16}$ m$^{-3}$, yet higher in the 600 A case, with relative maxima at R = 0 (centre of the machine) and R $\simeq$ 6 cm. The electron temperature is in the order of $\sim 5-10$ eV.

Comparing homologous discharges in presence and absence of SH (dotted and solid lines of the same colour respectively, in figure \ref{fig:exp_data}), the only relevant difference observed concerns the $n_e$ profiles measured by LP 5U, which display a remarkable decrease at values of the radial coordinate $\lesssim 3$ cm, i.e. at the centre of the device, when the SH is inserted. Indeed, LP 5U is situated \textit{behind} the SH, i.e. on the opposite side of the SH with respect to the positions of the gas puff valve and ECRH resonant surface (see figure \ref{fig:2D}): the suppression of $n_e$ profiles observed is thus reasonably due to the shadowing effect associated to the presence of the SH itself. For all plasma experiments, the values of neutral pressure measured by the two hot-cathode gauges at sections 3 and 5 of the device lie in the range of 0.09-0.1 Pa: this is essentially independent of the magnetic field configuration but is rather fixed by the volume of the vessel once the gas puff and pumping speed are set.

As concerns the setup of plasma simulations described in section \ref{sec:plasma_simulation}, the possible presence of the central SH is neglected and more details about this methodological choice are presented in subsection \ref{sub:simulation_setup}. On the contrary, the presence of the SH is included in the global erosion simulation, as it will be analysed in section \ref{sec:erosion}.

\section{SOLPS-ITER plasma simulation}
\label{sec:plasma_simulation}
The present section focuses on the numerical investigation of GyM plasma. In subsection \ref{sub:simulation_setup} the simulation setup and main modelling hypotheses are discussed, while the outcomes of the analysis are summarised in \ref{SOLPS_results}. The latter is further articulated into \ref{parametric_scan} about a parametric sensitivity scan, \ref{subsub:benchmark} in which the model is validated against experimental data and \ref{subsub:MS}, in which the effect of including He metastable states in SOLPS runs is assessed and compared with respect to a point model for GyM plasma \cite{tonello_point_2021}.

\subsection{Simulation setup}
\label{sub:simulation_setup}
\begin{figure} 
  \centering
  \includegraphics[width=0.7\textwidth]{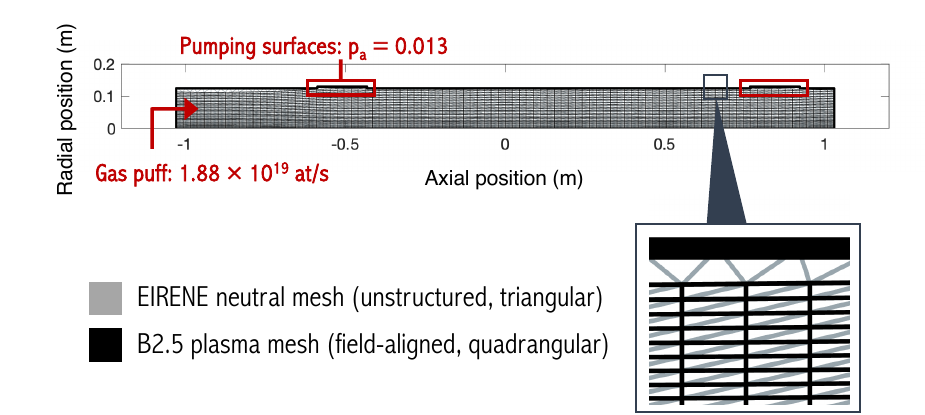}
  \caption{B2.5 and EIRENE computational mesh at $I_{coil}$ = 600 A plotted on the GyM vessel. Gas puff location and pumping surfaces are also displayed.}
  \label{fig:mesh}
\end{figure}

The first step towards the SOLPS-ITER plasma simulation is the construction of the computational mesh, which is done for linear machines starting from the reconstruction of magnetic equilibria for each discharge, according to the procedure described in detail in reference \cite{sala_simulations_2020}. The resulting B2.5 and EIRENE meshes, structured quadrangular and unstructured triangular respectively, are plotted in figure \ref{fig:mesh} for the plasma at $I_{coil}$ = 600 A. Since the magnetic configuration is almost purely axial for all cases, no significant differences in the appearance of the computational meshes are observed for the three different $I_{coil}$. 

The computational meshes displayed cover only one half of the physical domain and cylindrical symmetry around the axis of the machine is assumed. Also note that, while the EIRENE neutral mesh covers the entire volume of the device upon symmetrization, the B2.5 field-aligned plasma mesh extends between the two bases (targets) of the cylinder axially, and to the first point of tangency radially, i.e. not to the physical lateral wall of the device. \par

All plasma simulations include \ce{He+} and \ce{He++} as ion species and neutral ground state \ce{He}, the standard SOLPS reaction database for a pure He case is implemented. However, for the sole simulations discussed in subsection \ref{subsub:MS}, long-lived singlet and triplet He metastable states are also included among neutral species in addition to the ground state. For such runs, a wider reaction database is coherently implemented as described in more detail in \ref{subsub:MS}. \par

A gas puff strength of $1.88 \times 10^{19}$ neutral He atoms$/$s is set to match the experimental puffing conditions. All the wall surfaces but those corresponding to the turbomolecular pumps are considered as saturated with He atoms, their particle absorption probability is thus set as $p_a = 0$. On the contrary, a particle absorption probability at EIRENE pumping surfaces of $p_a = 0.013$ is set to mimic the experimental pumping speed\footnote{The particle absorption probability, or \textit{albedo}, at the pumping surfaces $p_a$ is set to model the effect of turbomolecolar pumps. Its value is related to the turbo-pump speed $S$ (Ls$^{-1}$) by: $S = A \times p_a \times 3.638 \times \sqrt{T/m}$, where $T$(K) is temperature and $m$(amu) is the mass of the neutral species. In the formula, $A$(cm$^2$) is the \textit{effective} area of the pumping surface: since axial symmetry is assumed in EIRENE, the area of each pumping surface to be considered for the computation of $p_a$ is $A = 2 \pi r_p L_p$, where $r_p$ and $L_p$ are the distance from the axis and the width of the pumping surfaces \cite{Owen2017}.}. Drift effects are totally neglected. The cross-field transport is modelled as diffusive, which requires the specification of ion diffusivity $D_n$ and ion/electron heat diffusivities $\chi_{i,e}$, treated as free input parameters.


\begin{figure} 
    \centering
    \begin{subfigure}[b]{\textwidth}
        \centering
        \includegraphics[width=\textwidth]{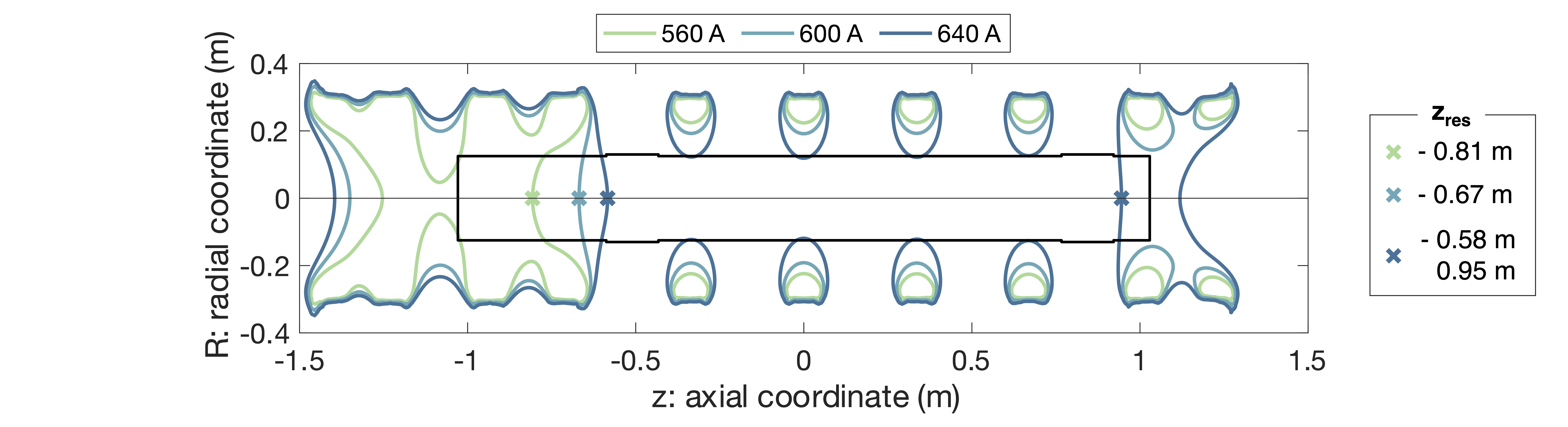}
        \caption{The solid coloured curves plotted on the GyM vessel represent the loci of points where the electron cyclotron resonance condition (i.e. B = 87.6 mT) is met for the three values of $I_{coil}$ considered. The crosses identify the points at which the curves intersect the cylinder axis, whose axial coordinates are named $z_{res}$ and listed in the legend. Note that in the $I_{coil}$ = 640 A case, two distinct points on the axis satisfy the resonance condition.}
        \label{fig:ECRH_loci}
    \end{subfigure}
    \begin{subfigure}[b]{\textwidth}
        \centering
        \includegraphics[width=\textwidth]{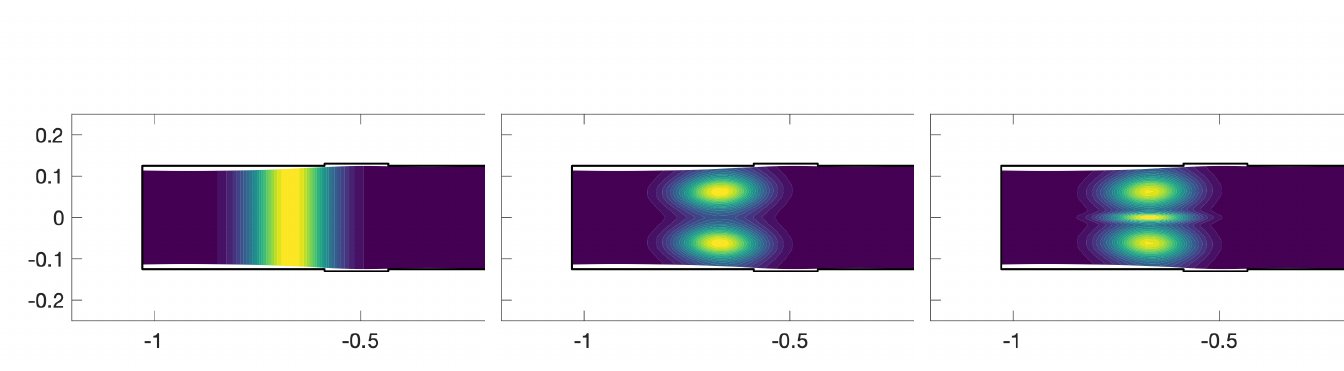}
        \caption{Contour plots of possible external power density source distributions implemented in this work to model the ECRH power absorption by the magnetized plasma. Although the resonance surface is not flat (i.e. the axial position of the resonance varies with the radial coordinate, see figure \ref{fig:ECRH_loci}), the axial profile of the external power source is modeled as a Gaussian curve centred at $z_{res}$ (taken as radially constant) with a fixed width $\sigma=0.1$ m. In turn, the amplitude of the Gaussian is modulated by a radial function, either constant or a with a variable number of Gaussian peaks.}
        \label{fig:ECRH_sources}
    \end{subfigure}
    \caption{ECRH external power source modelling in SOLPS-ITER simulations of GyM.}
    \label{fig:ECRH}
\end{figure}

Note that, depending on $I_{coil}$, the axial position of the resonant surface - i.e. the locus where the electron cyclotron frequency matches the gyrotron frequency of 2.45 GHz and energy is favourably absorbed by electrons, which happens for a magnetic field intensity of 87.5 mT - is displaced (see figure \ref{fig:ECRH_loci}). Therefore, the external power delivered to the plasma is modelled as an energy source for electrons distributed around the axial position $z_{res}$ of the ECRH resonance. \par

In all simulations, the power density axial profile is modelled as a Gaussian centered at $z_{res}$ with fixed width $\sigma=0.1$ m. Although the nominal power of the ECRH source is set, the efficiency in the absorption of energy by the electron population is unknown. Therefore, the actual amount of power $P_{ext}$ absorbed by the plasma (i.e. the 3D integral of the power density distribution) is treated as a free input parameter.

In the lack of experimental data or first-principle simulations of plasma-microwave interaction, the radial distribution of external power density source is also treated as a further degree of freedom. The modelling approach devised in this work includes either constant or Gaussian radial distributions, or a linear combination of the two (figure \ref{fig:ECRH_sources}). Again, any radial distribution of the transport coefficients or power source is specified only on one half of the computational domain (i.e. for $R>0$) and cylindrical symmetry is assumed.

As concerns the setup of boundary conditions, the sheath boundary conditions are enforced at the bases of the cylinder. On the symmetry axis of the cylinder, zero particle and energy fluxes, as well as zero parallel velocity gradient for the parallel momentum equation and zero current for the potential equation are imposed. On the lateral boundary, zero parallel velocity gradient and zero current are fixed, together with a \textit{leakage} condition prescribing particle and energy outflows suppressed by a factor $\alpha_{leak}=1 \times 10^{-3}$ with respect to the corresponding flows at the plasma sheath entrance. \par

Although some experimental setups described in section \ref{reference_experimental_scenario} include the presence of a sample-holder along the plasma stream for the exposure of selected samples for PMI experiments, the plasma simulations performed, described and validated against experimental data in the following sections do not account for its presence. \par

This choice is dictated by the limitations of the B2.5 mesh generator adopted and technical restrictions in SOLPS-ITER when modelling linear geometries, which require the specification of no more than two targets where the contact between the plasma and the solid boundary occurs. This constraint prevents the straightforward construction of a regular plasma computational mesh covering the entire volume of the device and including the sample holder. This is possible in principle, as demonstrated by the work of Rapp et al. \cite{Rapp2015}, where modifications to the SOLPS5.0 code were made to accommodate such a geometry, but is beyond the scope of the present paper. \par



A possible straight-forward alternative to the code modification would be to restrict the B2.5 computational mesh to the sole portion of the volume of the device between one of the two bases and one of the two faces of the SH including the gas puff valve and the ECRH resonant surface. However, this approach would exclude the portion \textit{behind} the SH, where - according to the experimental data discussed in \ref{reference_experimental_scenario} - the only significant differences with respect to the plain scenario are expected, and therefore it is disregarded. \par

In summary, as long as an overall plasma analysis is concerned, the presence of the SH is neglected. However, it will be taken into account properly in the following investigation of material erosion (section \ref{sec:erosion}).

\subsection{Results}
\label{SOLPS_results}

\subsubsection{Parametric scan} 
\label{parametric_scan}
\begin{figure} 
  \centering
  \includegraphics[width=\textwidth]{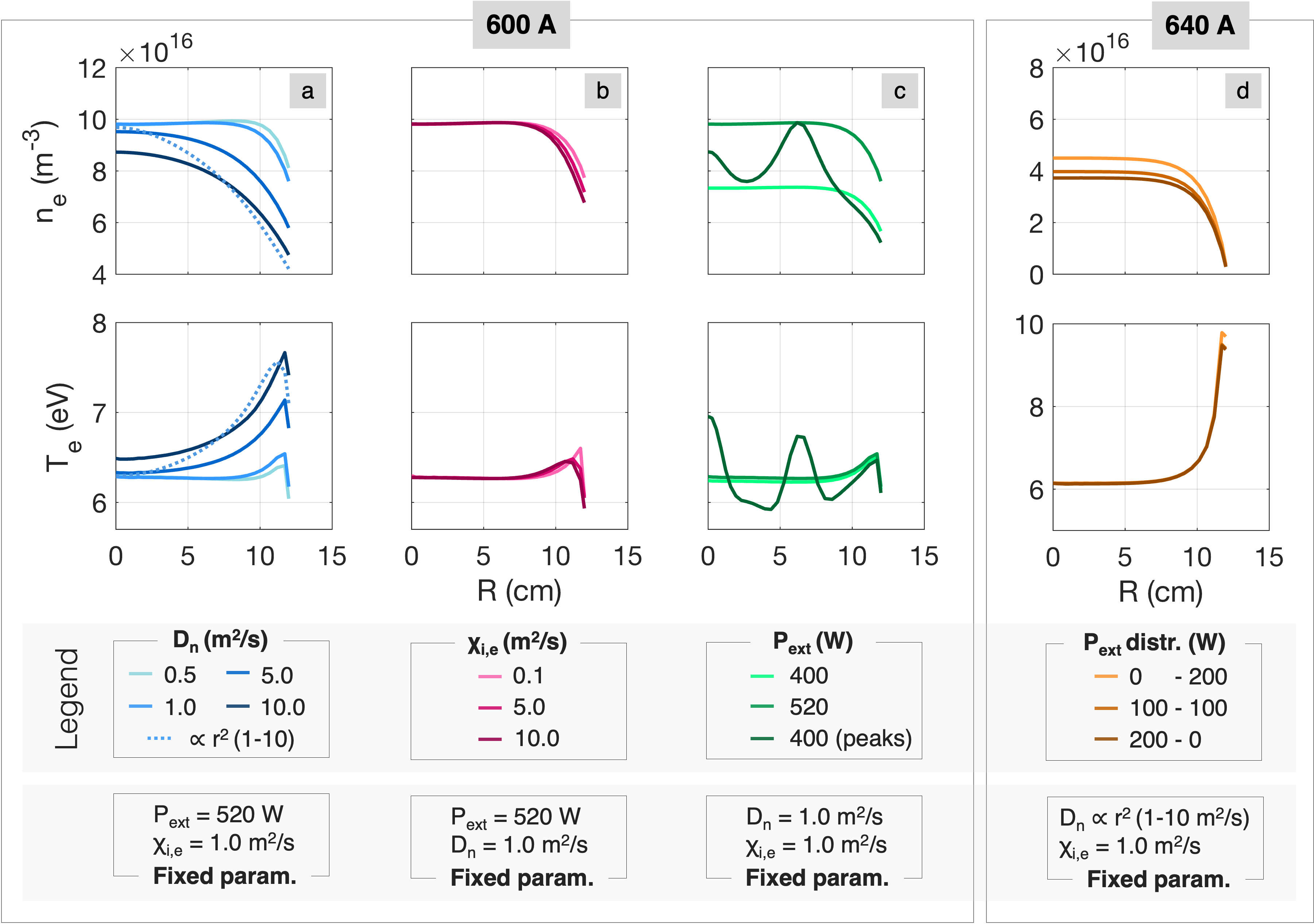}
  \caption{Sensitivity scan of one SOLPS-ITER free parameter at the time for GyM. For the $I_{coil}$ = 600 A case: a) particle diffusivity $D_n$, b) Electron/ion heat diffusivity $\chi_{i,e}$, c) total external power source $P_{ext}$. For the $I_{coil}$ = 640 A case: d) scan in the relative distribution of the same external total power source between the two resonance surfaces. }
  \label{fig:scan}
\end{figure}

As a first outcome of the numerical investigation of the He plasma in GyM, a sensitivity scan of the code's free parameters (enumerated in \ref{sub:simulation_setup}, i.e. particle and heat anomalous cross-field transport coefficients, external power source absorbed by the plasma) is presented. In each column of the left portion of figure \ref{fig:scan}, the impact of tuning one input parameter at the time is assessed for the $I_{coil}$ = 600 A scenario on the electron density and temperature radial profiles evaluated at the axial position of LP 4U (similar results hold for 560 A and 640 A scenarios as well). The values assumed by the parameter being scanned are reported in legend, while the values kept fixed for the other input parameters are listed at the bottom of each column. \par $\\$

\begin{enumerate}[(a)]
\item \textit{Particle diffusivity} $D_n$. An increase in the anomalous cross-field particle diffusion coefficient (figure \ref{fig:scan}a) reflects in a smoothing of the electron density gradient from the periphery of the device inward, and in a corresponding peaking of the electron temperature profile for $D_n \lesssim$ 5 m$^2/$s. For $D_n \gtrsim$ 5 m$^2/$s, an increase in $D_n$ translates the profiles to lower densities and higher temperatures respectively at constant shape. In figure, the outcome of a simulation with $D_n \propto r^2$ ranging between 1 m$^2/$s and 10 m$^2/$s - which will be used for achieving a better agreement with experimental data in section \ref{subsub:benchmark} - is also shown.

\item \textit{Heat diffusivities} $\chi$. A change in the ion and electron heat diffusivities (figure \ref{fig:scan}b) has negligible impact over both electron and temperature profiles, which is consistent with the fact that GyM plasma conditions - characterised by low densities and small parallel temperature gradients - fall into the sheath-limited regime, dominated by convection rather than conduction \cite{uccello_linear_2023}. \par

\item \textit{External power source} $P_{ext}$. An increase in the strength of the external power source (figure \ref{fig:scan}c) results in an overall increase of the electron density profile at constant electron temperature. This is coherent with the GyM plasma being a low density one: the extra power supplied is spent to increase the plasma ionization fraction rather than to heat up electrons. Moreover, at fixed total amount of external power supplied, the modulation of the radial distribution of the power density source - according to a linear combination of constant and Gaussian functions - results in a local increase of electron density and temperature in correspondence of the peaks of the distribution.
\end{enumerate}

According to figure \ref{fig:ECRH_loci}, for the sole 640 A coil current scenario, two resonant surfaces along the cylinder axis - and two corresponding values of $z_{res}$ - may be identified in sectors 2 and 6 of the device (with reference to figure \ref{fig:2D}): in principle, the absorption of external energy occurs at both axial positions. The right portion of figure \ref{fig:scan}d compares the electron density and temperature profiles yielded by three simulations for which the same amount of input total power (i.e. $P_{ext}$ = 200 W) is split in different ratios between the two resonant surfaces (0-200 W, 100-100 W, 200-0 W, where the first number is the amount of external power deposited at sector 2 and the second at sector 6). \par

The temperature profiles are coincident, and only limited difference in the magnitude of electron density profiles at constant shape is observed: $n_e$ evaluated at the axial position of LP 4U increases as the amount of external power deposited at the second resonance in sector 6 increases from 0 to 200 W. This is true not only at the axial position of LP 4U, but at all axial positions. \par

For simplicity and similarity with the other coil current scenarios, in the continuation of the analysis, the total amount of power - treated as a totally free parameter - will be modelled as absorbed only at the resonance in sector 2, i.e. on the gas puff side of the device.


\subsubsection{Validation}
\label{subsub:benchmark}
In the present subsection, a validation of the numerical results against experimental data is proposed. This is meant (i) to provide a realistic plasma background for the eventual ERO2.0 erosion analysis, (ii) to gain physical insight on the possible mechanisms responsible for the observed experimental profiles, mostly in relation to the different magnetic field configurations taken into account, (iii) to complement the dataset and enrich the characterisation of GyM He plasma.

The parametric scan presented in subsection \ref{parametric_scan} pointed out that, among the plasma parameters for which experimental data are available, electron density profiles are the most sensitive to a change in the input parameters. On the contrary, the values of electron temperature - especially at the centre of the device - are scarcely dependent on the free parameters and are rather fixed by the throughput strength, i.e. the overall balance of gas puffing and pumping. \par

Operatively, an optimum set of input parameters (cross-field transport coefficients, external power absorbed by the plasma and radial distribution of the external power density) is identified for each $I_{coil}$ scenario to provide a satisfactory qualitative and quantitative reproduction of the experimental density profiles, and the temperature ones are obtained accordingly. 

In all three cases, a parabolic profile is imposed for particle diffusivity (i.e. $D_n \propto r^2$), such that $D_n(R=0)\sim 1$ m$^2/$s and $D_n(R=R_0)\sim 10$ m$^2/$s, at fixed and radially uniform heat diffusivity $\chi_{i,e}=1$ m$^2/$ s. The magnitude of the external power input required to match the experimental profiles for the three coil current intensity cases needs to be different ($P_{ext} = $ 200 W, 300 W and 250 W respectively). Also, seeking for a satisfactory agreement with the density profiles - which all show relative maxima at R = 0 and R $\simeq 6$ cm (see \ref{reference_experimental_scenario}) - a radial distribution for the external power density source is introduced (namely a linear combination of a constant and two Gaussian functions in the form: $\rho(r)= N \times \left\{ c_0 + \sum_{i=1}^2 c_i \exp \left [- \left (r-r_i \right )^2/\sigma_i^2 \right ] \right\}$, with $c_0$, $c_i$, $r_i$, $\sigma_i$ free parameters and $N$ normalization factor to ensure that the volume integral of the power density distribution matches the desired external power $P_{ext}$). Physically, this suggests that the efficiency of ECRH-plasma coupling and energy transfer is largely dependent on the specific magnetic field configuration, which in turn can promote the absorption of energy at selected radial spots, where higher electron density is found. In the absence of specific experimental data, a first-principle numerical simulation of wave propagation in the GyM magnetized plasma by means of $ad-hoc$ multiphysics codes could provide further physical insight, but this is beyond the scope of this paper. \par

\begin{figure} 
  \centering
  \includegraphics[width=\textwidth]{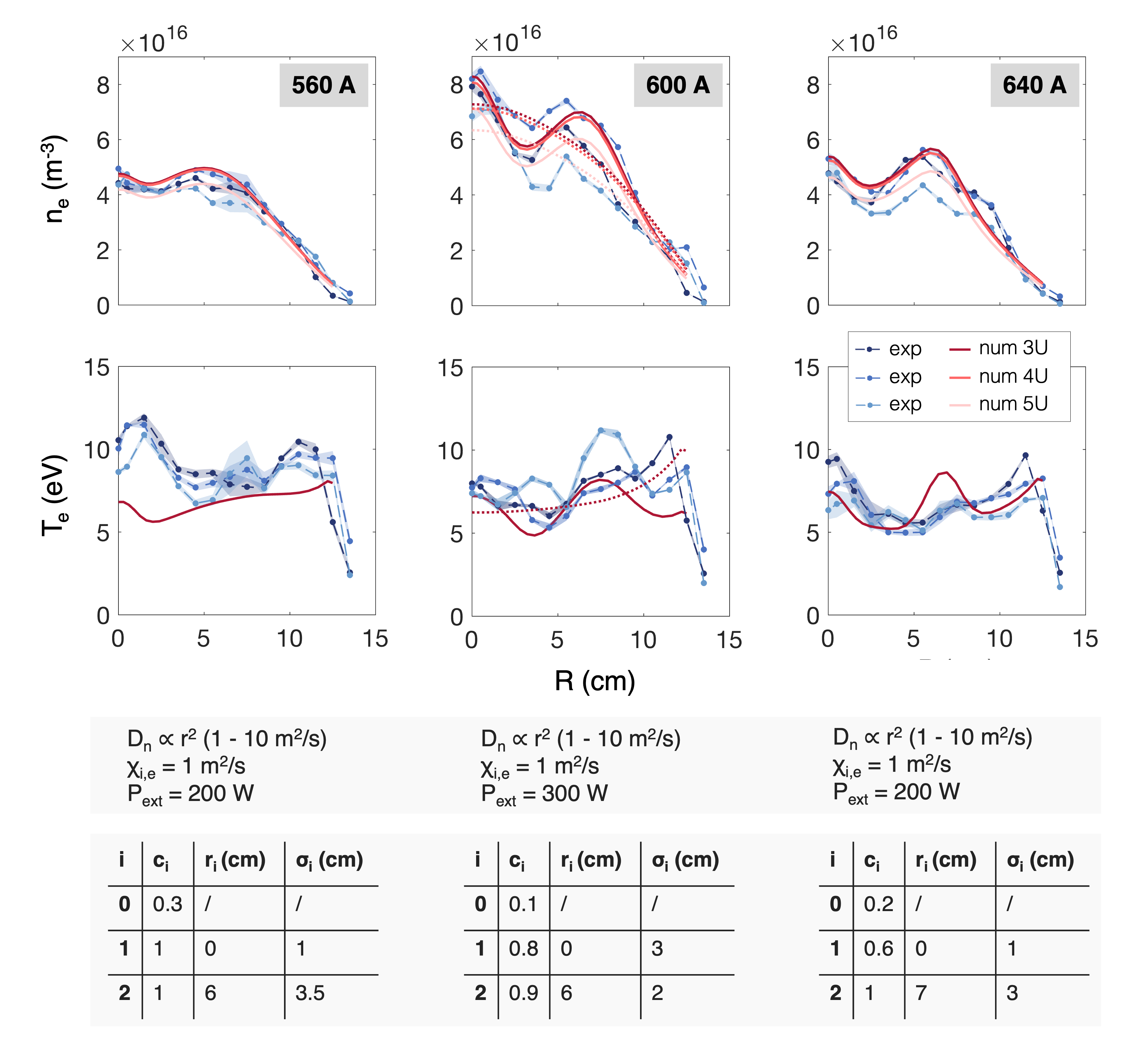}
  \caption{Validation of SOLPS-ITER simulation results against electron density and temperature experimental profiles discussed in section \ref{reference_experimental_scenario}. Blue and red lines represent the experimental and numerical radial profiles respectively. Each tone of blue indicates one of the three LP and is associated to a corresponding tone of red, which indicates the axial position of the LP itself at which the numerical profiles are taken. At the bottom of the figure, the first horizontal band summarises the free parameters (anomalous cross-field transport coefficients and external power source) adopted for the numerical simulations presented, while the second horizontal band provides the parameters used to model the radial distribution of the power source $\rho(r)$ according to the formula $\rho(r)= N \times \left\{ c_0 + \sum_{i=1}^2 c_i \exp \left [- \left (r-r_i \right )^2/\sigma_i^2 \right ] \right\}$. For the 600 A case, the results of simulations at constant radial power deposition - used for the erosion analysis presented in section \ref{sec:erosion} - are plotted in dotted line style.}
  \label{fig:benchmark}
\end{figure}

The results and a summary of the adopted parameters are visible in figure \ref{fig:benchmark}. The numerical temperature profiles are coincident at the three LP axial positions; however, while they are in good agreement with the corresponding experimental ones for the 640 A scenario, they tend to underestimate them in the 600 A and even more in the 560 A scenario. \par

For the 600 A case, the dotted line in figure \ref{fig:benchmark} shows the reference plasma background used for the erosion analysis presented in section \ref{sec:erosion}. This simulation uses the same anomalous transport coefficients enforced in the solid-line simulations ($D_n \propto r^2$ in the range of 1-10 m$^2/$s, $\chi_{i,e} = 1.0$ m$^2/$s), with a radially uniform external electron energy source of 400 W. The numerical results well reproduce the observed radial $n_e$ and $T_e$ overall trends, although the density peak at $R \simeq 6$ cm is not reproduced. However, note that the radial portion of plasma relevant to the ERO2.0 erosion study is the one striking the SH, which extends well below the radial position of the $n_e$ peak. Therefore, these simplified numerical profiles will constitute the plasma background for the ERO2.0 erosion simulation discussed in section \ref{sec:erosion}.



By considering the latter simplified 600 A model with some detail from now on (similar considerations hold for the other scenarios as well), the 2D distributions of neutral \ce{He}, \ce{He+}, \ce{He++} densities and of neutral He pressure are plotted on a longitudinal section of the device. The density of neutral He is about three orders of magnitude greater than the density of \ce{He+}, which is in turn five orders of magnitude greater than the one of \ce{He++}, which determine an ionization fraction $\alpha \simeq 0.002$. The presence of \ce{He++} may be thus reasonably neglected in the low-density and low-temperature GyM plasma, contrary to the situation typical of tokamaks, where the higher ionization state is generally dominant \cite{tonello_phd}. \par

The neutral pressure distribution is approximately uniform throughout the central volume of the device and is peaked at the bases of the cylinder, where the ions striking the solid surfaces are recycled as neutrals and puffing is performed. Also note that the He pressure provided by the model in correspondence of the pressure gauges is about 0.09 Pa: this value is totally compatible with the experimental one (see \ref{reference_experimental_scenario}).

\begin{figure} 
  \centering
  \includegraphics[width=\textwidth]{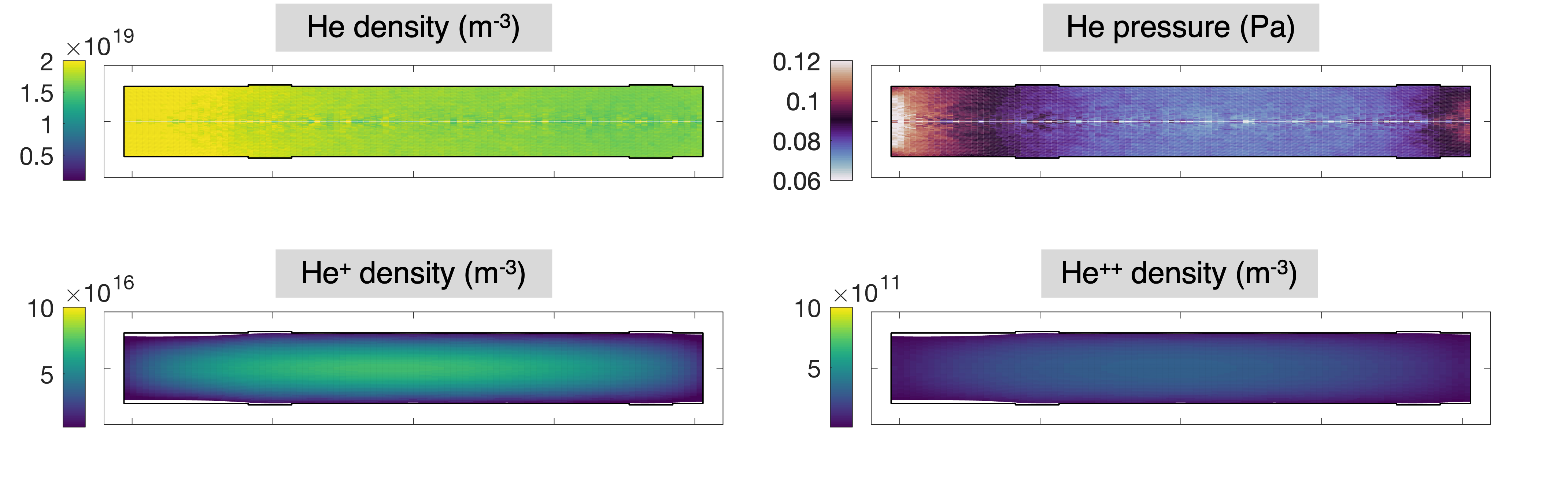}
  \caption{2D distribution of \ce{He}, \ce{He+}, \ce{He++} density and \ce{He} pressure plotted on a longitudinal section of GyM for the $I_{coil}$ = 600 A scenario. Note that $n_{\ce{He}} \gg n_{\ce{He+}} \gg n_{\ce{He++}}$.}
  \label{fig:2D_map}
\end{figure}

\subsubsection{Role of metastable states}
\label{subsub:MS}
The He atom possesses two long-lived metastable (MS) states: both feature the $1s^{1}2s^{1}$ electronic configuration, and are distinguished by the spin orientation of the two electrons into singlet and triplet states, which will be labeled as \ce{He^{*(1)}} (with lifetime $\tau \simeq$ 20 ms) and \ce{He^{*(3)}} ($\tau \simeq$ 7800 s) respectively \cite{Sun2020}.

In low-temperature, low-density plasma conditions typical of LPDs, MS states and their transport within the chamber could affect the plasma properties. In this work, the possible relevance of including such He MS states among the plasma neutral populations is preliminarily investigated for the first time in a purely He plasma by means of \textit{ad-hoc} SOLPS-ITER simulations. Operationally, this goal is achieved by splitting the neutral population treated by EIRENE into three separate species (the ground state \ce{He} and the two MS states) and enlarging the reaction database to include the MS-resolved ones extracted from the AMJUEL database and reported in table \ref{MS_reaction} \cite{amjuel}.\par

\begin{table}[]
\centering
\begin{tabular}{l|l|l|l} \toprule
    {} & {Reactions} & {Type} & {Database} \\ \midrule
    1  &  \ce{e- + He -> 2e- + He+}             & EI & AMJUEL 2.3.9b  \\ 
    2  &  \ce{e- + He       -> e- + He^{*(1)}}  & EI & AMJUEL 2.3.9c  \\
    3  &  \ce{e- + He       -> e- + He^{*(3)}}  & EI & AMJUEL 2.3.9d  \\
    4  &  \ce{e- + He^{*(1)} -> e- + He}        & EI & AMJUEL 2.3.9e  \\
    5  &  \ce{e- + He^{*(1)} -> 2e- + He+}      & EI & AMJUEL 2.3.9f  \\
    6  &  \ce{e- + He^{*(1)} -> e- + He^{*(3)}} & EI & AMJUEL 2.3.9g  \\
    7  &  \ce{e- + He^{*(3)} -> e- + He}        & EI & AMJUEL 2.3.9h  \\
    8  &  \ce{e- + He^{*(3)} -> e- + He^{*(1)}} & EI & AMJUEL 2.3.9i  \\
    9  &  \ce{e- + He^{*(3)} -> 2e- + He+}      & EI & AMJUEL 2.3.9j  \\
    10 &  \ce{e- + He+ -> He}                   & RC & AMJUEL 2.3.13b \\
    11 &  \ce{e- + He+ -> He^{*(1)}}            & RC & AMJUEL 2.3.13c \\
    12 &  \ce{e- + He+ -> He^{*(3)}}            & RC & AMJUEL 2.3.13d \\
    \bottomrule
\end{tabular}
\caption{List of He MS-resolved atomic reactions included in SOLPS-ITER model, extracted from AMJUEL database. EI and RC stand for electron impact collision and recombination respectively \cite{amjuel}.}
\label{MS_reaction}
\end{table}

\begin{figure} 
  \centering
  \includegraphics[width=0.75\textwidth]{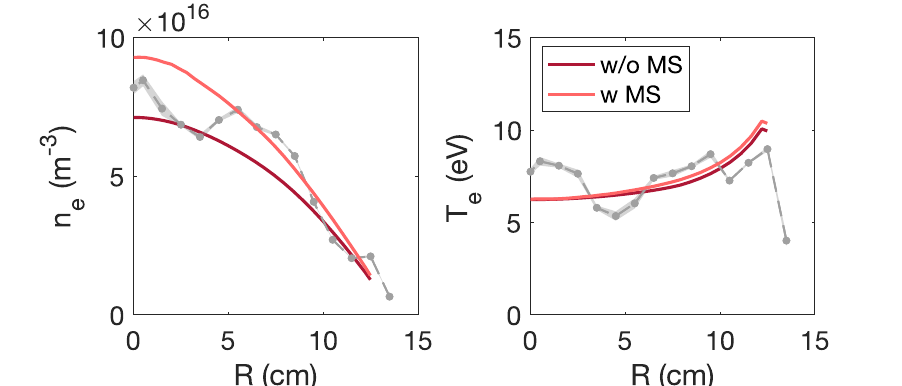}
  \caption{Electron density and temperature radial profiles yielded by homologous MS-resolved and unresolved simulations based on the simplified $I_{coil}$ = 600 A case described in \ref{subsub:benchmark} and evaluated at the axial position of the LP 4U. In background, experimental data acquired by LP 4U (600 A) are showed for reference.}
  \label{fig:MS_comp}
\end{figure}

Figure \ref{fig:MS_comp} compares the electron density and temperature radial profiles at a fixed axial position yielded by two simulations, MS-unresolved and resolved respectively. The input parameters set in the two cases are those of the simplified model for the $I_{coil}=$ 600 A case presented in \ref{subsub:benchmark} (dotted red lines in figure \ref{fig:benchmark}: $D_n \propto r^2$ in the range of 1-10 m$^2/$s, $\chi_{i,e} = 1.0$ m$^2/$s, $P_{ext} = 400$ W uniformly distributed radially). The inclusion of MS neutral states yields a slightly higher electron density profile, acting (check the power scan in \ref{parametric_scan}) as if a larger amount of power was supplied to the plasma: the presence of He MS species thus appears to promote ionization processes and increase the overall ionization fraction. Conversely, the two electron temperature profiles are practically indistinguishable. Also, the distribution of both MS species provided by Eirene is approximately uniform throughout the chamber. \par

Nevertheless, in the absence of a proper spectroscopic diagnostics, the actual relevance of He MS states in GyM plasma goes undetected under the experimental point of view, and the resolved model may not be validated. As a partial compensation for such limitation, a comparison between the outcomes of the SOLPS-ITER runs of figure \ref{fig:MS_comp} and the results of a simplified MS-resolved 0D plasma model is proposed. Such 0D model was developed in \cite{tonello_phd} extending the MS-unresolved counterpart firstly presented in \cite{tonello_point_2021}, and its main features are retraced in Appendix A. \par

In the 0D model, the experimental values of gas puff and pumping speed (42 sccm and 500 Ls$^{-1}$ respectively) are set, a decay length for density of $\lambda_n = 5$ cm is set coherently with the choice made in reference \cite{tonello_point_2021}. Concerning the cross-field diffusion coefficient, the SOLPS simulations validated in section \ref{subsub:benchmark} all assumed a radially variable profile. In a 0D treatment, any spatial dependence is suppressed, and the specification of a single value of diffusivity $D_{\perp}$ is required. Therefore, a scan in the value of the diffusion coefficient is made investigating the effect of its choice on the outcomes. \par

Clearly, the strongly different hypotheses on which SOLPS and the 0D model are based make it impossible to perform a direct quantitative benchmark between the two. Most importantly, the 0D model includes one MS species only and is based on the ADAS database, while SOLPS treats both MS states and is based on the AMJUEL database. However, the global 0D model is a useful simplified tool to compare relative trends between MS-resolved and unresolved cases. \par

\begin{figure} 
  \centering
  \includegraphics[width=0.57\textwidth]{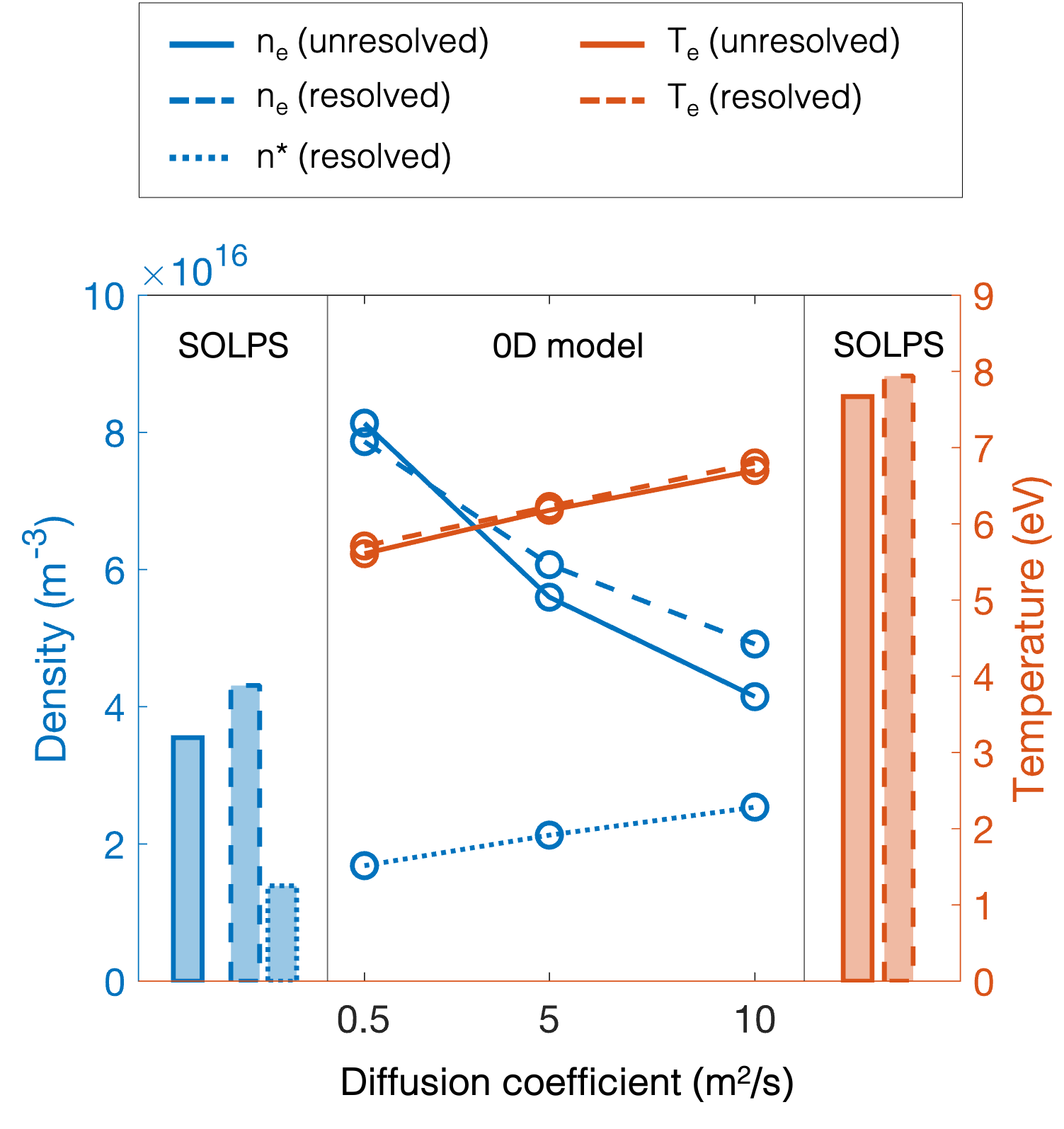}
  \caption{Comparison between the results of SOLPS-ITER and 0D point model as concerns MS-unresolved and resolved He plasma simulations. The bar charts are related to SOLPS results: on the left, the mean electron density in the MS-unresolved and resolved case (solid and dashed respectively) and the mean density for MS species (dotted) are reported. On the right, the mean electron temperature is reported for the unresolved (solid) and resolved (dotted) simulations. In the centre, the same quantities provided by the 0D model described in Appendix A are shown with the same line styles as a function of the diffusion coefficient.}
  \label{fig:MS_0D_SOLPS}
\end{figure}

The results of the comparison are summarised in figure \ref{fig:MS_0D_SOLPS}. The bars report the values of the mean metastable and electron density and temperature for the two SOLPS simulations - MS-unresolved and resolved respectively - whose outcomes are plotted in figure \ref{fig:MS_comp} according to the legend. Conversely, the line plot in the centre shows the same quantities as foreseen by the 0D point model as a function of the diffusion coefficient $D_{\perp}$ input parameter. \par

As a first observation, both the SOLPS simulation and the 0D model - independently on the $D_{\perp}$ set - predict a negligible increase in the electron temperature $T_e$ when turning from the unresolved to the resolved case. Also, the ratio between the mean MS density and the mean electron density foreseen by the 0D model increases from $\sim 20 \%$ to $\sim 50 \%$ as $D_{\perp}$ increases from 0.5 to 10 m$^2$/s: the ratio at $D_{\perp}$ = 5 m$^2$/s is $\sim 30 \%$ and is consistent with the one predicted by the SOLPS simulation.

Turning from the unresolved to the resolved model in SOLPS foresees an increase in the mean $n_e$ of $\sim 20 \%$. As far as the point model is concerned, the magnitude and the sign of the variation in $n_e$ following the introduction of the MS population appears to depend on $D_{\perp}$. For a choice of $D_{\perp}$ = 10 m$^2$/s, an increase of $\sim 20 \%$ in $n_e$ is observed, which is compatible with SOLPS results -. However, lower values of $D_{\perp}$ are associated to a decrease of $n_e$ upon introduction of MS species.


\renewcommand{\arraystretch}{1.2}

Overall, although no rigorous validation may be provided at this stage, one may argue that the current implementation of He MS populations in SOLPS-ITER is at least partially supported by the compatibility of its results with the ones of a much simpler 0D model.

Although assessing the role of MS species in a pure He plasma is crucial for the characterisation of the plasma itself and a validation of the MS-resolved model is urgent, in the following erosion analysis we will continue to refer to the unresolved simulations presented in subsection \ref{subsub:benchmark}, which were validated against experimental data. In the first instance, it appears reasonable to assume that the actual role of MS states in GyM plasma is scarcely relevant as long as the numerical investigation of PMI is concerned. In fact, since erosion results from the impact of energetic ions on solid surfaces, and since the external power source is also treated as a free parameter, once the numerical electron density profiles match the experimental ones, it makes little difference that this is achieved without MS states among the neutral population and slightly higher input power or with MS states and slightly lower input power respectively. \par

\section{ERO2.0 global erosion simulation}
\label{sec:erosion}
The present section focuses on the material side, i.e. it presents the ERO2.0 simulations performed to assess the erosion of plasma-facing structures in GyM, of PMI-relevant tungsten (W)-based samples exposed to the plasma stream and the deriving deposition. The main goals, in this respect, are (i) the comparison between the outcome of simulations based on simplified assumptions for the wall material with respect to the full steel composition, (ii) the assessment of the effect related to SH presence on wall erosion and, conversely, (iii) the role of wall impurities on sample erosion. Note that, although no experimental data concerning material erosion are available for model validation at this stage, the results of this work will guide the design of an experiment devoted to the collection of erosion-deposition data, to be performed by placing collectors at selected spots of the vessel.


\subsection{Simulation setup}
\begin{figure} 
    \centering
    \begin{subfigure}[b]{\textwidth}
        \centering
        \includegraphics[width=0.63\textwidth]{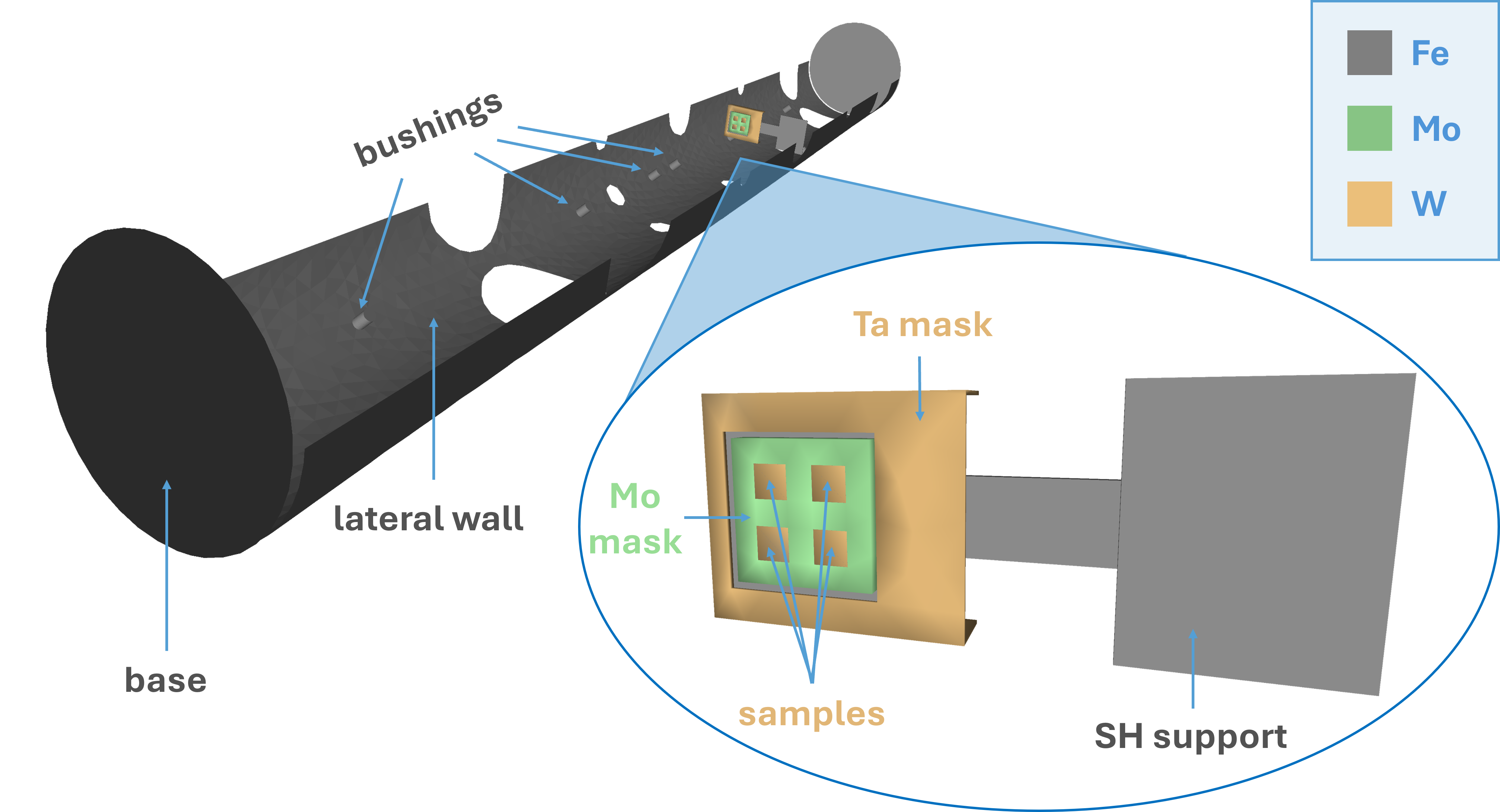}
        \caption{All 3D components of GyM drawn and imported in ERO2.0, with a zoom on the sample-holder. Only half of the lateral wall with relative bushings on it is represented for better visibility. Colours represent the materials adopted in ERO2.0 simulation, as reported in legend. \\}
        \label{fig:EROgeom}
    \end{subfigure}
    \begin{subfigure}[b]{\textwidth}
        \centering
        \includegraphics[width=0.43\textwidth]{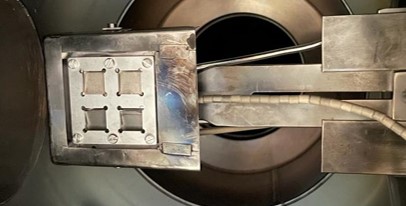}
        \caption{Photograph of GyM sample-holder.}
        \label{fig:SHphoto}
    \end{subfigure}
    \caption{3D GyM geometry considered in this work.}
    \label{fig:geom}
\end{figure}

As far as the simulation setup is concerned, the 3D GyM wall geometry is reproduced (figure \ref{fig:geom}) and the 2D distribution of relevant plasma parameters (electron and ion density and temperature, plasma flow) yielded by the simplified SOLPS-ITER simulation at $I_{coil}$ = 600 A - taken as a reference and discussed in detail in the second part of \ref{subsub:benchmark} - is imported as a plasma background, assuming cylindrical symmetry. Note that, according to the observations made in \ref{subsub:benchmark}, \ce{He+} may be safely considered as the only relevant ion species, thus excluding the \ce{He++} contribution from the input plasma background. Moreover, the aforementioned sample-holder is included in the definition of the simulation domain at fixed plasma background, figure \ref{fig:geom} showing its geometry and materials. In all simulations, sputtering yields are retrieved from the SDTrimSP database available in ERO2.0. The energy distribution of sputtered species is modelled as a Thompson one, while their angular distribution is assumed to be a function of the incidence angle of the impinging ion, with the forward direction with respect to it being favoured \cite{yamamura1981}.  \par

A first parameter which has been varied in the simulations is the wall material. Although the GyM vessel is made of AISI 304L SS, the previous study treated it as made of either pure Fe or pure Cu \cite{alberti_global_2023}. Such a choice was dictated by a lack of sputtering yield data for He ion species on chromium (Cr) and nickel (Ni), i.e. the main alloying elements of SS, and their similarity - in terms of physical properties - to Cu, for which a much wider database was available and thus taken as proxy \cite{alberti_global_2023}. Since then, the dataset has been enlarged by means of SDTrimSP simulations, and this work considers and compares pure Fe, pure Cu and realistic AISI 304L SS (73\% Fe, 18\% Cr, 9\% Ni) as possible wall materials: the results of this investigation are presented in \ref{subsub:wallMaterial}. \par

Secondly, the possibility of setting a bias voltage ($V_{bias}$) to the SH is foreseen. Applying a tunable bias voltage to the SH enables to modulate the kinetic energy of the plasma ions impinging onto it. The application of a bias voltage in experiments influences the conditions in the sheath. However, since SOLPS-ITER background plasma only extends up to the sheath entrance, no influence of the bias voltage on the plasma parameters imported in ERO has been considered. For $V_{bias} = 0$, the energy of impinging ions depends solely on the plasma potential, which is about 20 V. Although the bias voltage applied is always negative (thus accelerating impinging ions), we will always refer to its absolute value for simplicity. \par

The $V_{bias}$ values applied to SH in this simulations reflect the ones usually set experimentally, in the range 0 - 320 V. As in experiments, the bias voltage is applied only to samples and molybdenum (Mo) mask. Also note that the tantalum (Ta) mask, being unbiased during experiments, is expected not to be eroded, and it has been assumed as made of W due to a lack of sputtering yield data for all material combinations with Ta and to reduce the number of different species in the simulations. The choice of W is justified by the high sputtering threshold for He on Ta ($\sim$ 90 eV), which is comparable with the one on W (105-110 eV) \cite{eckstein1993}.  \par

\subsection{ERO2.0 results}

\subsubsection{Wall material analysis}
\label{subsub:wallMaterial}
\begin{figure} 
  \centering
  \includegraphics[width=0.64\textwidth]{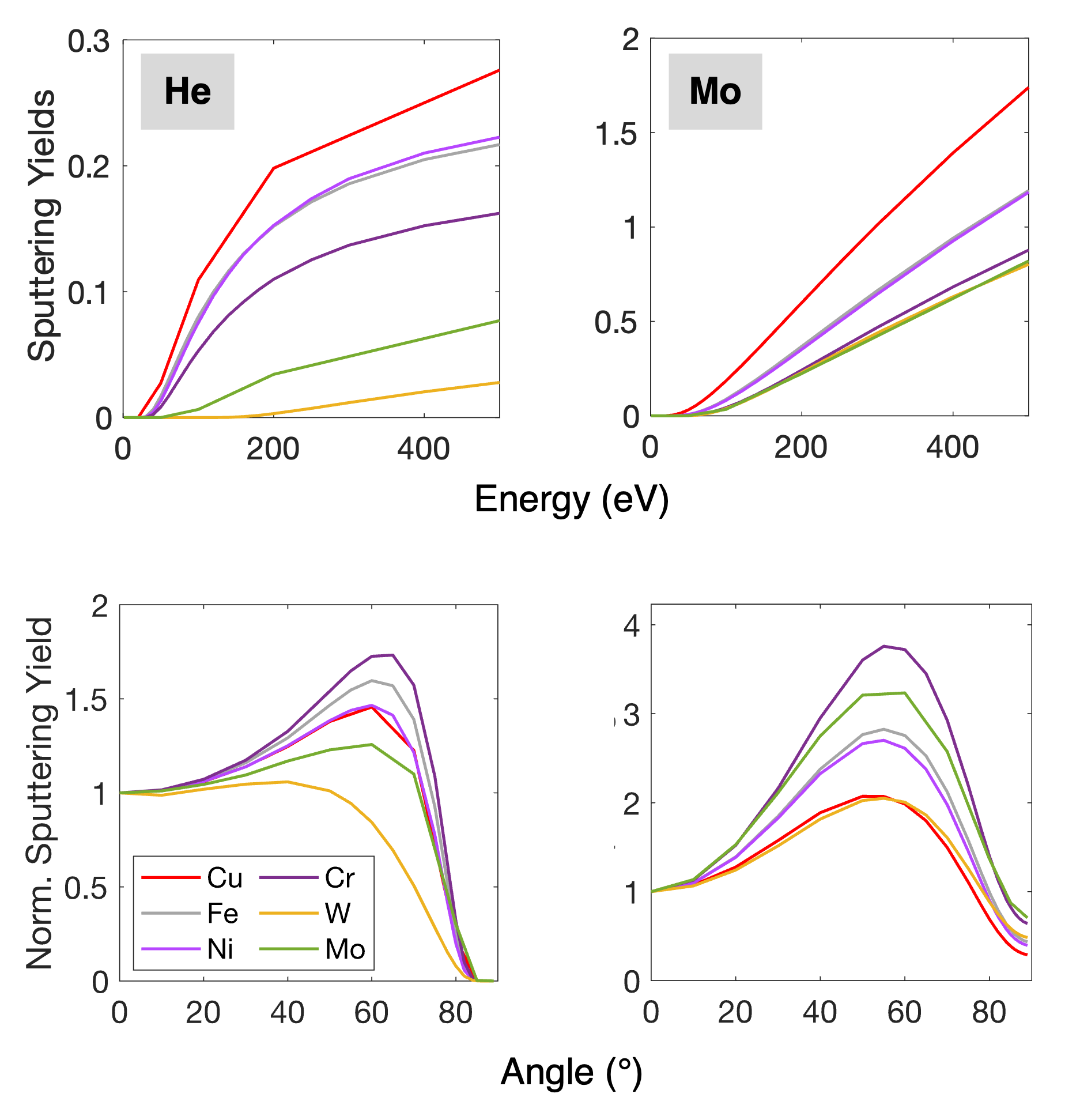}
  \caption{Sputtering yields for He (left) and Mo (right) projectile on different target materials as a function of energy at orthogonal incidence (first row) and of impinging angle at 200 eV (second row, normalized to 0° value).}
  \label{fig:yields}
\end{figure}

In this section, the effect of wall material choice on GyM wall erosion is investigated. Three different wall material compositions have been considered: (i) pure Cu, (ii) pure Fe and (iii) the real AISI 304L SS. For each one of these, five different values of $V_{bias}$ have been set, namely 0, 50, 120, 200 and 320 V, in agreement with the ones commonly adopted in experiments. The dependence of the SDTrimSP sputtering yields used in this work on projectile energy and incidence angle is reported in figure \ref{fig:yields}. Due to the numerous combinations of projectile - target materials (36), only the most relevant ones are shown, namely He, as main plasma species, and Mo, as main eroded species, projectiles. For the SS case, according to the local incoming flux of \ce{He+} ions and eroded particles, the code evaluates the sputtering yield for each element separately, and the eroded flux is then weighted on the material composition. No composition evolution is considered in this first attempt. \par

The estimated gross erosion of GyM walls, i.e. the number of sputtered atoms per unit area and time without considering their possible deposition, as a function of wall material with no bias voltage applied to the SH and expressed in $[$atoms $\cdot$ m$^{-2} $s$^{-1}]$ is: (i) 7.92$\times 10^{17}$ for pure Cu, (ii) 5.28$\times 10^{16}$ for pure Fe, and (iii) 3.86$\times 10^{16}$ for SS. \par

For the case of SS wall, the overall gross erosion arises from the sum of the gross erosion individual contributions associated to each element in the SS composition. Pure Cu wall presents a severely higher gross erosion with respect to both pure Fe and SS, whose behaviour is comparable. This hierarchy is in accordance with the relative magnitude of the sputtering yields reported in figure \ref{fig:yields}, where Cu presents the highest erosion for both He and Mo projectiles. Note that the observed trend is unchanged also at a larger $V_{bias}$, for which a few percent increase in the global gross erosion amount is obtained for all wall materials investigated: this is consistent with the higher kinetic energy acquired by plasma ions impinging on the SH, which reflects in a higher amount of energy retained by Mo and W atoms eroded from the SH itself, eventually impinging on surrounding walls and fostering their erosion. \par

The overall 30\% increase in gross erosion for Fe with respect to SS walls may be ascribed to the presence of Cr in SS, which is characterized by the lowest sputtering yield. Indeed, about 80\% of SS gross erosion is due to Fe, with a 10\% contribution due to Cr and Ni each: the presence of Fe will thus reduce as the fluence increases and erosion proceeds, in favour of a Cr-enriched surface. This could further reduce the overall SS gross erosion. Considering GyM vessel as entirely made of of Fe is thus expected to yield a reasonable upper limit for the wall erosion, while limiting the complication in the interpretation of the simulation results deriving from the large number of projectile-target combinations for SS. Therefore, this modelling approach will be adopted in the continuation of this work.



\subsubsection{Effect of sample-holder on GyM wall erosion}
In the previous work \cite{alberti_global_2023}, a proof of concept of the coupling between SOLPS-ITER and ERO2.0 in a linear device (GyM) was presented. In this work, the objective is moving forward in the direction of modelling a real PMI experiment, thus including also the SH in ERO2.0 simulation domain. \par

Due to its position in the middle of the device, impurities eroded from the SH are likely to impinge onto the lateral wall, which in turn may undergo erosion. Note that the magnetic field lines are parallel to the lateral wall itself: according to the leakage boundary conditions enforced in the setup of plasma simulations, the ion radial flux at the lateral boundary of the plasma computational mesh is 1/1000 of the corresponding one at the entrance of the sheath which builds up in front of a solid surface orthogonal to the magnetic field lines. Therefore, the lateral wall erosion is mainly due to impurities eroded elsewhere, with the SH possibly playing a major role as it is the only biased component during GyM experiments. All the other parts of GyM wall are not substantially influenced by the presence of the SH, thus will not be considered in this section: as a reference, at maximum $V_{bias}$, impurities eroded from the samples and SH are responsible for just $\sim 1 \%$ of the gross erosion of the cylinder bases, which may be practically fully ascribed to the striking of plasma ions. \par

\begin{figure} 
  \centering
  \includegraphics[width=0.7\textwidth]{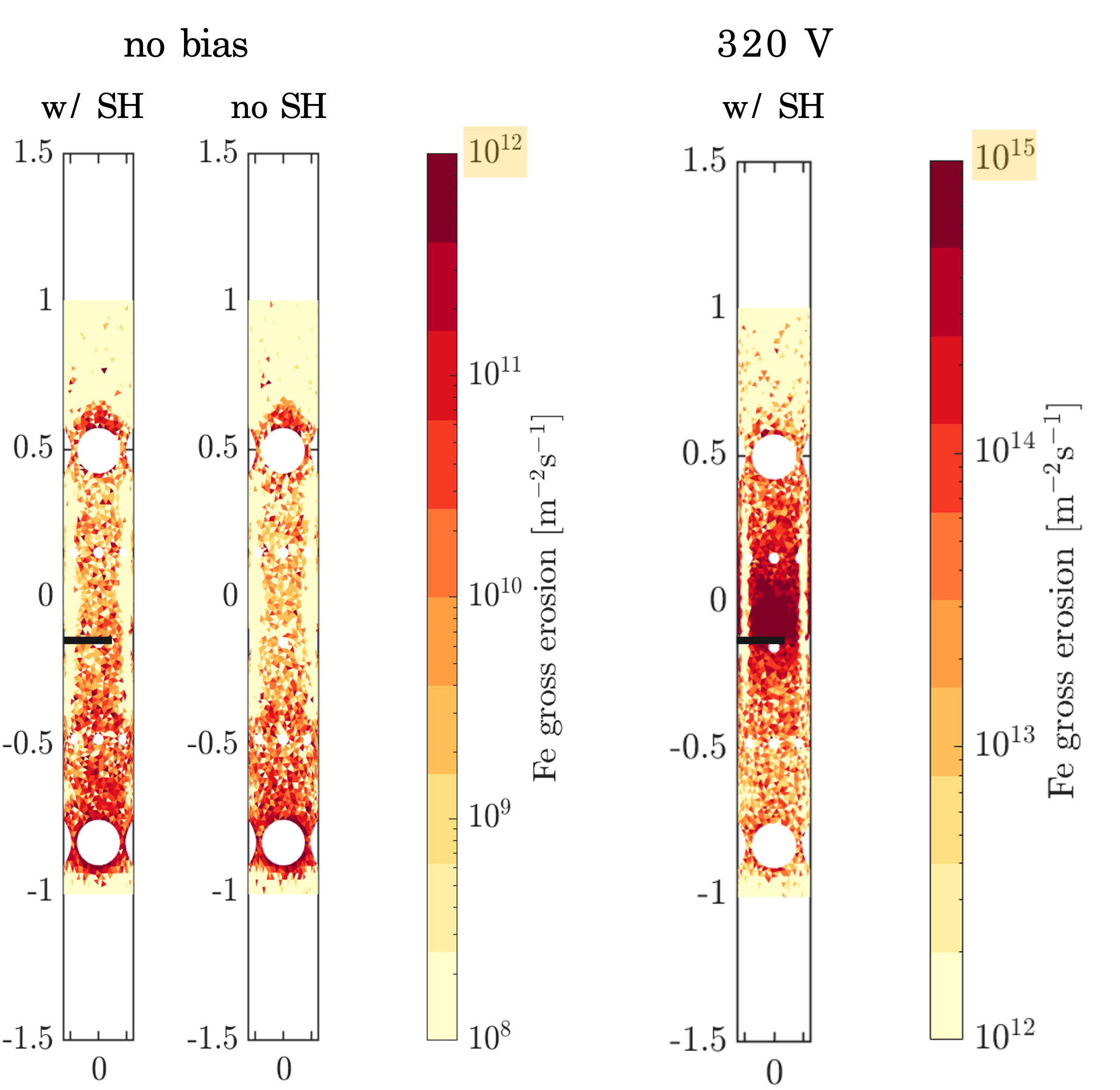}
  \caption{Gross erosion distribution of GyM lateral wall in different conditions, namely without SH inserted and including it with lowest (0 V) and highest (320 V) $V_{bias}$. Note the different boundaries for the colorbars in the two cases.}
  \label{fig:LatWallErosion}
\end{figure}
\begin{figure} 
  \centering
        \subfloat{
        \includegraphics[width=0.55\textwidth]{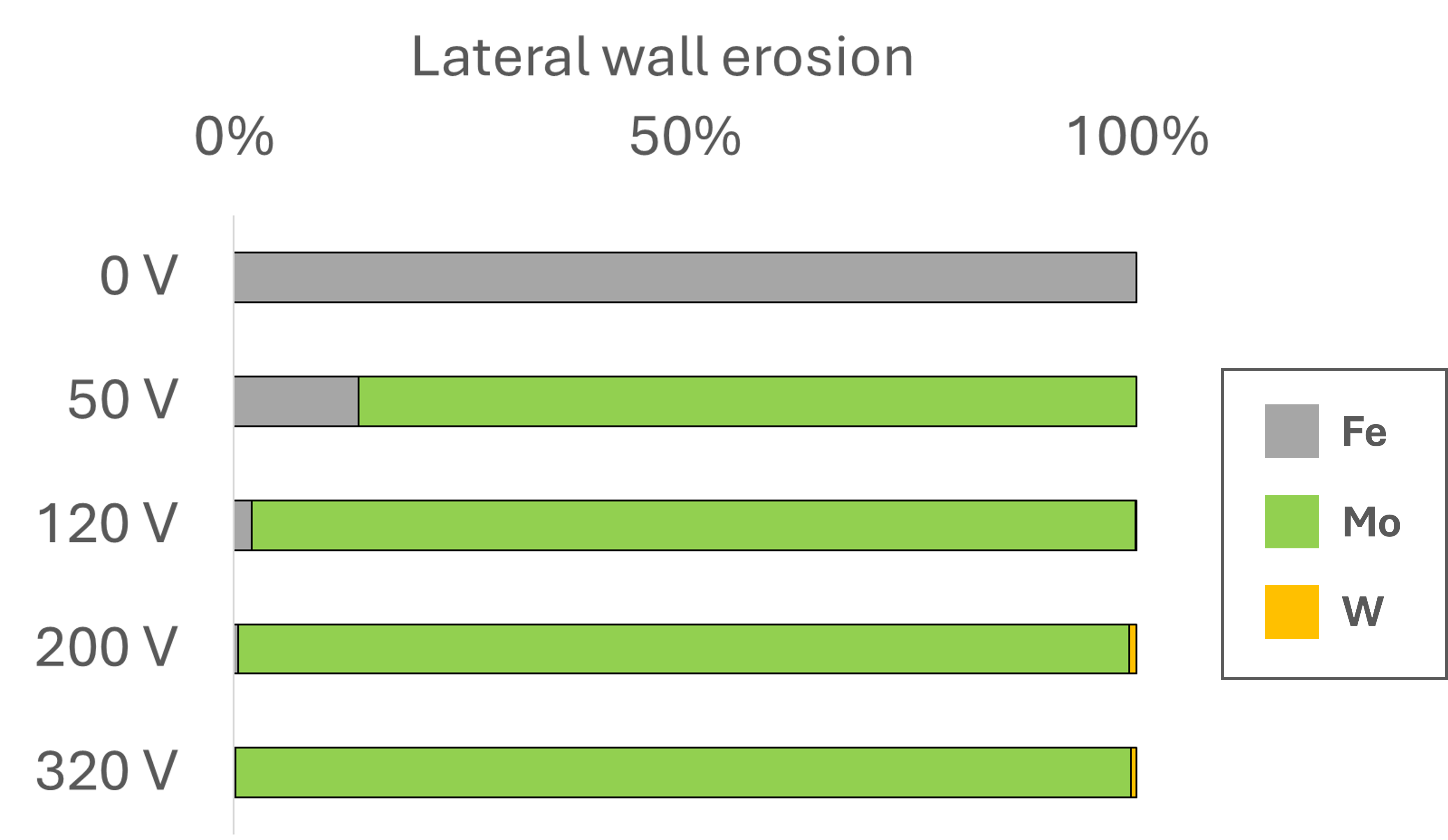}}
        \subfloat{
        \begin{tabular} {c|c}
        \multirow{2}{4em}{$V_{bias}$ (V)} & \multirow{2}{7em}{\centering Gross erosion (m$^{-2}$ s$^{-1}$)} \\ \\
        \hline 
        0 & $1.68 \times 10^{11}$ \\
        50 & $1.20 \times 10^{12}$ \\
        120 & $8.46 \times 10^{12}$ \\
        200 & $4.47 \times 10^{13}$ \\
        320 & $2.04 \times 10^{14}$ \\
        \end{tabular}}
    \caption{Contribution of the different eroded impurities to the erosion of lateral wall as a function of the bias voltage applied to the SH. He plasma is not included since it is not causing erosion of lateral wall. The overall gross erosion of the lateral wall is reported in table.}
    \label{fig:LatWallErosionFraction}
\end{figure}

Figure \ref{fig:LatWallErosion} shows the distribution of the gross erosion of GyM lateral wall in different conditions, namely without SH and with SH inserted, at lowest (0 V) and highest (320 V) $V_{bias}$. Complementarily, figure \ref{fig:LatWallErosionFraction} presents the relative contributions of the different species (i.e. eroded impurities) to the erosion of the lateral wall as a function of the bias voltage applied to the SH. Again, He plasma ions are not included since they are not striking the lateral wall and causing its direct erosion. \par

For $V_{bias}$ = 0, the influence of the SH on lateral wall erosion is limited and concentrated in the proximity of the SH itself. In this condition, only Fe from the SH support and cylinder bases is eroded and may in turn erode the lateral wall, while Mo and W are not: the contribution from the Fe SH support (see figure \ref{fig:EROgeom}) is then what differentiates the case with and without SH in figure \ref{fig:LatWallErosion}. As $V_{bias}$ is increased to 50 V, the kinetic energy of plasma ions overcomes the sputtering threshold for Mo. Eroded Mo impurities in turn retain sufficient energy to largely dominate the erosion of the lateral wall and increase its absolute value by one order of magnitude (figure \ref{fig:LatWallErosionFraction}). At higher and higher $V_{bias}$, the magnitude of the lateral wall erosion increases more and more and is firmly dominated by Mo. W eroded from samples plays a marginal role only for V$_{\text{bias}} >$ 120 V, when the sputtering threshold for He on W is overcome. According to the colorbars in figure \ref{fig:LatWallErosion} and the table in figure \ref{fig:LatWallErosionFraction}, the difference in both local and global gross erosion between the lowest and highest $V_{bias}$ reaches a few orders of magnitude.

\subsubsection{Effect of eroded materials on sample erosion}
Once investigated the effect of SH inclusion on the erosion of GyM walls, the influence of wall impurities on the sample erosion is assessed. This is particularly relevant for the possible interpretation of experimental data, since both the increased erosion caused by wall impurities and their deposition on samples could influence experimental outcomes of mass loss measurements. \par

\begin{figure} 
  \centering
        \subfloat{
        \includegraphics[width=0.55\textwidth]{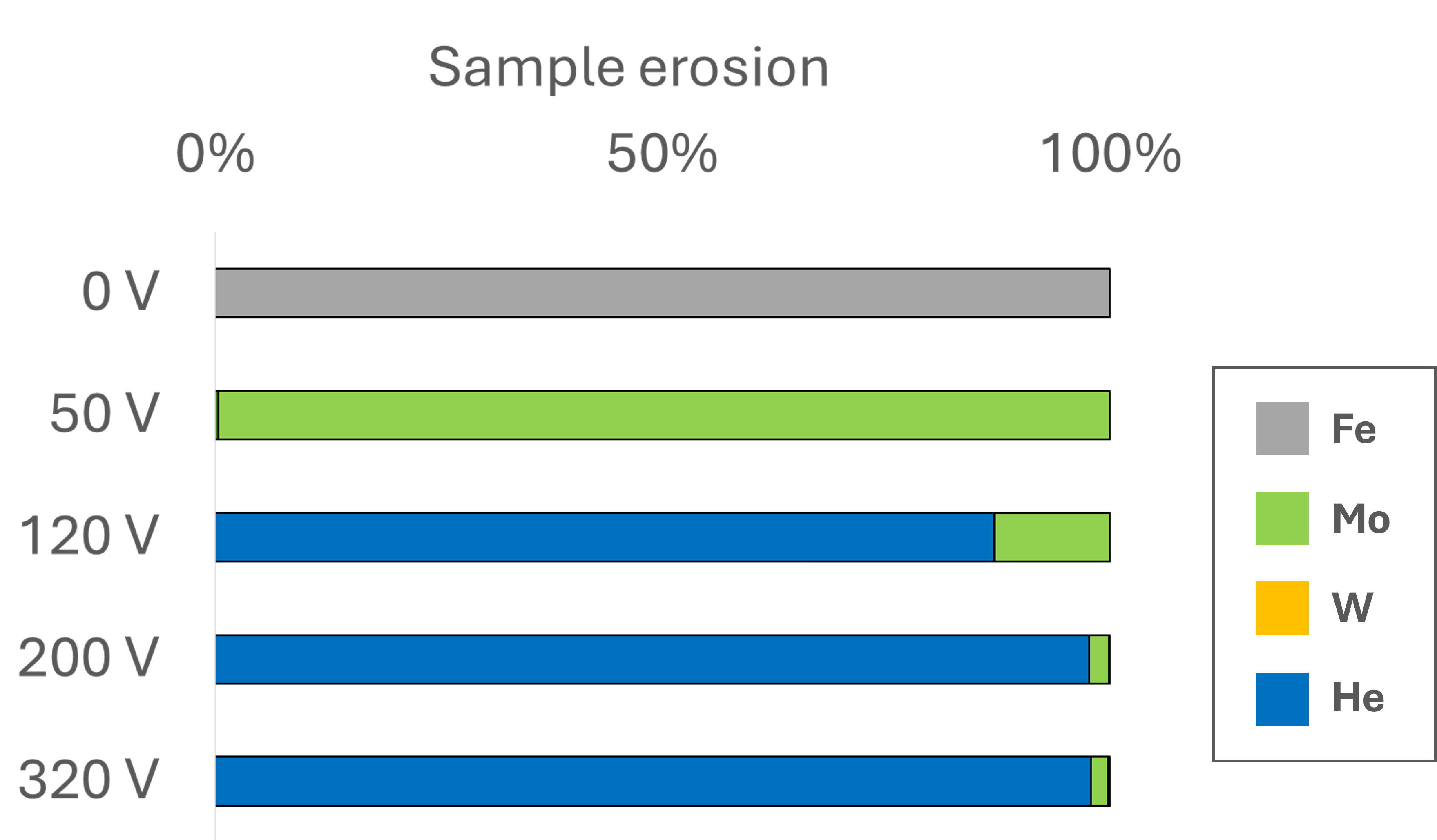}}
        \subfloat{
        \begin{tabular} {c|c}
        \multirow{2}{4em}{$V_{bias}$ (V)} & \multirow{2}{7em}{\centering Gross erosion (m$^{-2}$ s$^{-1}$)} \\ \\
        \hline 
        0 & $1.94 \times 10^{10}$ \\
        50 & $1.19 \times 10^{13}$ \\
        120 & $5.04 \times 10^{16}$ \\
        200 & $1.76 \times 10^{18}$ \\
        320 & $6.01 \times 10^{18}$ \\
        \end{tabular}}
    \caption{Contribution of the different impinging projectiles to the erosion of samples as a function of bias voltage applied to the SH. The overall gross erosion of the samples is reported in table.}
    \label{fig:SamplesErosionFraction}
\end{figure}

Figure \ref{fig:SamplesErosionFraction} shows the contribution of the various impinging particles to the gross erosion of samples as a function of the bias voltage applied to SH. In this case, He plasma is included, since samples are directly eroded by it. As for lateral wall, only Fe components are eroded by He plasma with no bias, thus only Fe neutrals and ions possess enough energy to slightly contribute to sample erosion. For $V_{bias}$ = 50 V, Mo mask starts to be eroded by the plasma, and it immediately becomes the major contributor to sample erosion. Indeed, both Mo eroded from the mask and W eroded from samples may be ionized in plasma, be captured by axial magnetic field and come back to sample-holder due to plasma flow. \par

As soon as He plasma ions, whose flux onto the samples is $\Gamma \simeq 3.5 \times 10^{20}$ m$^{-2} $s$^{-1}$, are able to directly erode W samples (for $V_{bias} \geq$ 120 V), they dominate sample erosion: this is consistent with the fact that the eroded impurities account for just $\sim 0.3 \%$ of the total particle flux striking the samples, which is thus largely dominated by the plasma itself. However, it should be noted that around the sputtering threshold for He on W, i.e. for $V_{bias}$ = 120 V, a 15-20\% of W sample erosion is still caused by impurities eroded from Mo mask. The contribution due to eroded impurities then decreases to few percent at higher bias in favour of He. Also note that, at every step in the $V_{bias}$, the sample gross erosion increases by orders of magnitude. \par

\begin{figure} 
  \centering
        \subfloat{
        \includegraphics[width=0.6\textwidth]{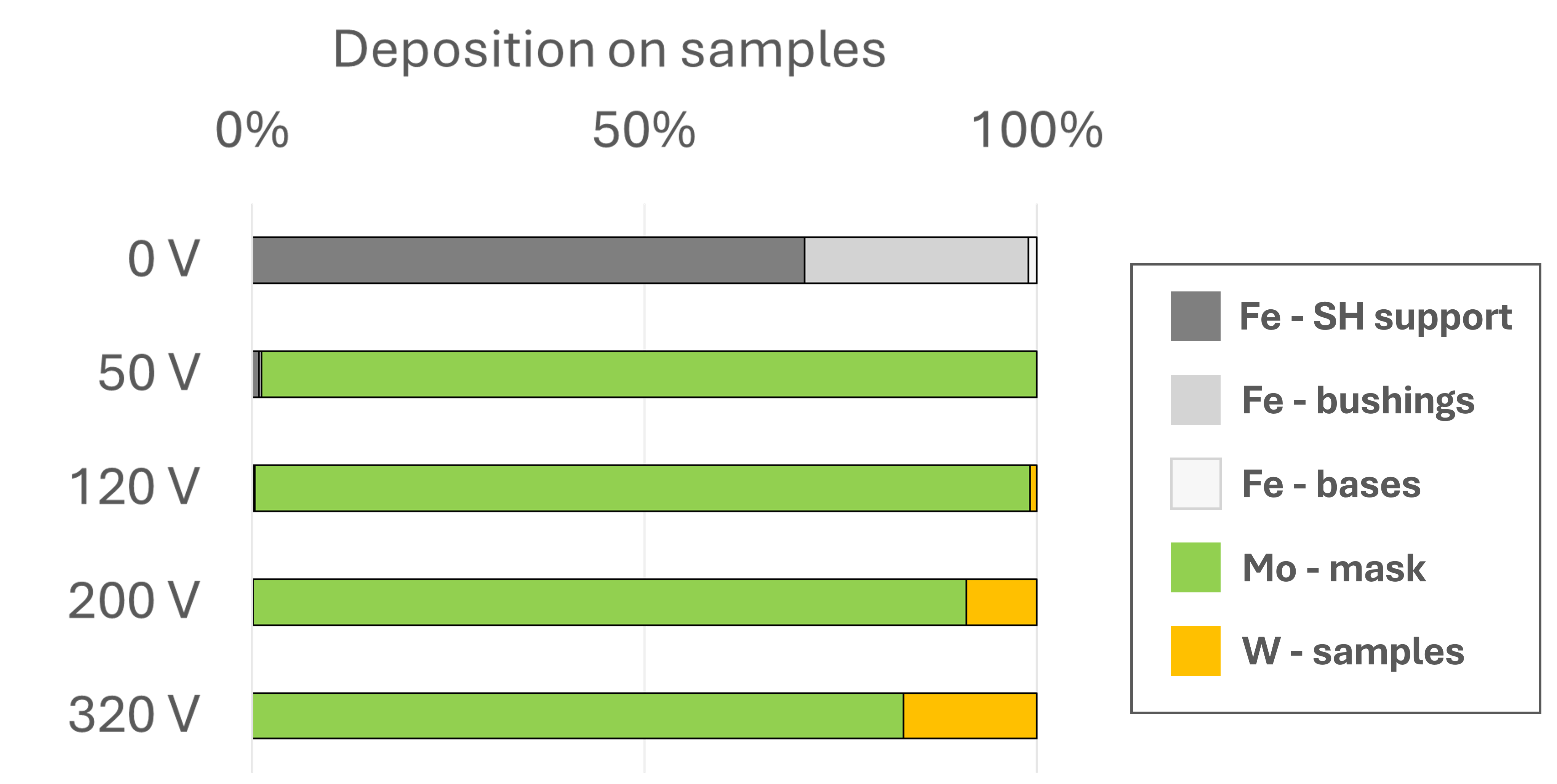}}
        \subfloat{
        \begin{tabular} {c|c}
        \multirow{2}{4em}{$V_{bias}$ (V)} & \multirow{2}{7em}{\centering Deposition (m$^{-2}$ s$^{-1}$)} \\ \\
        \hline 
        0   & $5.60 \times 10^{11}$ \\
        50  & $3.96 \times 10^{13}$ \\
        120 & $1.61 \times 10^{14}$ \\
        200 & $3.38 \times 10^{14}$ \\
        320 & $5.62 \times 10^{14}$ \\
        \end{tabular}}
    \caption{Deposition fraction of impurities on samples as a function of their erosion location, for all the $V_{bias}$ investigated. The total deposition on samples is reported in table.}
    \label{fig:SamplesDepositionFraction}
\end{figure}

Finally, the fractional deposition of impurities on samples is evaluated in figure \ref{fig:SamplesDepositionFraction}. For each $V_{bias}$ investigated, the origin (i.e. the erosion location within the vessel) of the deposited species is displayed. For $V_{bias}$ = 0, the main source of deposition is Fe eroded from the SH support, with a non-negligible contribution due to Fe eroded from bushings \cite{alberti_global_2023} around the SH (see figure \ref{fig:EROgeom}). As it is evident from the first row of figure \ref{fig:SamplesDepositionFraction}, the bases of the GyM wall are too far to significantly influence sample erosion/deposition. As $V_{bias}$ increases, Mo coming from the mask plays a dominant role even at the highest $V_{bias}$, when its contribution remains higher than redeposition from samples themselves. Looking at table in figure \ref{fig:SamplesDepositionFraction}, it should be noted that deposition on samples is well below the gross erosion for $V_{bias} \geq$ 120 V, thus only marginally influencing net erosion measurements when bias is high enough for He ions to cause W sputtering. On the contrary, for $V_{bias} <$ 120 V, samples are in net deposition conditions. 

\section{Conclusions and future perspectives}
\label{sec:conclusion}
The study discussed in this paper presents a comprehensive numerical investigation of Plasma-Material Interaction (PMI) in GyM linear plasma device, performed by coupling plasma solver SOLPS-ITER and erosion-deposition code ERO2.0, according to the procedure firstly described in reference \cite{alberti_global_2023}. To this purpose, few PMI-relevant pure He discharges are taken as a reference experimental framework for plasma background production. \par

Based on a preliminary sensitivity scan, which points out that the tuning of SOLPS-ITER free parameters mostly affects the electron density rather than temperature radial profiles, a validation of the numerical plasma profiles against available experimental data is presented for three magnetic field configurations. Optimal input parameters allowing to reproduce experimental Langmuir probe profiles are identified, allowing (i) to recognise the strength and radial shape of the external power source as the major responsible for the difference in the profiles at different coil current intensity, (ii) to generate a realistic plasma background (at $I_{coil}$ = 600 A) taken as a reference for eventual erosion investigation. \par

Secondly, a novel investigation of the role of neutral long-lived He metastable states is carried out through proper SOLPS simulations. Primarily, the inclusion of MS populations in GyM conditions results in an overall increase of the electron density at constant external power source, which is coherent with the results yielded by the analytical global point model presented in Appendix A of this work for a suitable choice of diffusion coefficient. \par

The outcomes of the reference MS-unresolved SOLPS run are taken as input for ERO2.0 studies of impurity erosion and deposition concerning both the walls of the device and the samples exposed to the plasma stream and arranged in a central sample-holder (SH), as a function of the wall material and of the bias voltage applied to the SH itself. With respect to the preliminary work presented in \cite{alberti_global_2023}, the present investigation is extended to account for the realistic SS composition of GyM vessel on one hand, and for the presence of the SH in the geometry on the other. \par

The estimation of the gross erosion as a function of the wall material points out that SS behaviour is well approximated by that of Fe only, which is then taken as proxy for the sake of simplicity. Moreover, the SH - mainly through the impurities eroded from the Mo mask - turns out to dominate the erosion of the lateral wall of the device. \par

As far as W-based samples exposed to the plasma stream are concerned, their erosion by \ce{He+} species is the dominant mechanism for sufficient bias voltage, although the Mo impurities eroded from the SH also contribute to the erosion of the samples themselves. Finally, the analysis of the impurities deposited on the samples suggests that they mainly originate from the Mo mask as the bias voltage exceeds the threshold for Mo sputtering by impinging plasma ions. \par

Several continuations and future perspectives are envisaged for the present work. On the plasma side, this includes the application of the new Wide Grid SOLPS-ITER release to GyM geometry, possibly enabling to build a computational mesh accounting for the presence of the SH. Also, the investigation of the role of He metastable states is only at the beginning: for instance, further numerical campaigns and more refined theoretical models are required to assess their contributions in tokamaks, possibly supported by experimental data for validation and comparison purposes. \par

Still on the plasma side, the first-principle modelling of wave transport in plasma through proper multiphysics code may shed light on the absorption of external power by electron cyclotron resonance at specific spots of the resonant surface, which in this work is taken as a purely free parameter and adapted to match the experimental profiles. \par

On the material side, an experimental campaign is foreseen to fill the current lack of data concerning the erosion and deposition of both GyM vessel and samples. The experiment, whose design is supported by the ERO2.0 numerical simulations, will be carried out by placing deposition monitors at selected spots of the device - where higher erosion and deposition rates are expected - and collected data will in turn allow to validate the outcomes of the numerical simulations presented in this paper.

A further possible continuation of the ERO2.0 numerical investigation is about assessing the effect of residual carbon or oxygen present in the GyM vacuum chamber on the erosion and deposition processes. Finally, the implementation of a back-coupling scheme between the two codes is of interest, such that the ERO2.0 impurity sources are provided as input to second-order SOLPS-ITER simulations. \par

\newpage

\appendix
\section{MS-resolved 0D model for GyM plasma}
In this appendix, some details about the MS-resolved 0D model used in section \ref{subsub:MS} are reported. Such analytical model is developed in \cite{tonello_phd} extending the MS-unresolved one presented in reference \cite{tonello_point_2021}, which is based on the volume averaging of SOLPS-ITER equations. \par

GyM plasma is described by a set of space-independent variables, i.e. density $n_p$ and temperature $T_p$ of each one of the considered populations $p$. The MS-unresolved model of reference \cite{tonello_point_2021} included ground state (GS) He as the only neutral species and \ce{He+} as the only ion species, thus neglecting the presence of \ce{He++} ions and allowing to identify the ion and electron density $n_{\ce{He+}}=n_e$ to ensure neutrality. Moreover, the unresolved reaction rates were obtained by solving the ADAS radiative collisional model neglecting the lifetime of MS states. \par

Conversely, a resolved treatment explicitly introduces a density balance equation for every MS species considered. For simplicity, the 0D model developed in \cite{tonello_phd} and hereby retraced still considers \ce{He+} as the only ion species and a single MS state He$^{*}$ accounting for both the singlet and triplet states, which thus requires the introduction of a single additional density balance equation. Ion and neutral temperatures are assumed to be constant and equal to the room temperature of 0.025 eV. The reaction rates are extracted from the corresponding ADAS MS-resolved database. \par

The equations of the resolved model are reported below, where the unknowns $n_{\ce{He+}} = n_e$ are the densities of \ce{He+} ions and electrons respectively, $n_{\ce{He}}$ and $n^{*}$ are the densities of the ground state neutral He and of the single MS species respectively, and $T_e$ is the electron temperature. 

\begin{equation}
    \frac{dn_{\ce{He+}}}{dt} = R_{iz} n_e n_{\ce{He}} + R^{*}_{iz} n_e n^{*} - R_{rc} n_e n_{\ce{He+}} - R^{*}_{rc} n_e n_{\ce{He+}} - \Gamma_{i,\ce{wall}} n_{\ce{He+}}
    \label{E1}
\end{equation}

\begin{equation}
\begin{split}
    \frac{dn_{\ce{He}}}{dt}& = -R_{iz} n_e n_{\ce{He}} + R_{rc} n_e n_{\ce{He+}} + X_{\ce{(MS)\rightarrow (GS)}} n_e n^{*} \\ 
    & - X_{\ce{(GS)\rightarrow (MS)}} n_e n_{\ce{He}} + \Gamma_{n,\ce{recyc}} n_{\ce{He+}} - \Gamma_{n,\ce{pump}} n_{\ce{He}} + \frac{\Gamma_{n,\ce{puff}}}{V}
\end{split} 
\label{E2}
\end{equation}

\begin{equation}
\begin{split}
    \frac{dn^{*}}{dt}& = -R^{*}_{iz} n_e n^{*} + R^{*}_{rc} n_e n_{\ce{He+}} + X_{\ce{(GS)\rightarrow (MS)}} n_e n_{\ce{He}} \\
    & - X_{\ce{(MS)\rightarrow (GS)}} n_e n^{*} - \Gamma_{n,\ce{pump}} n^{*} 
\end{split} 
\label{E3}
\end{equation}

\begin{equation}
\begin{split}
    \frac{3}{2} n_e \frac{dT_e}{dt}& = \frac{P_{ext}}{V_e} - E_{iz} R_{iz} n_e n_{\ce{He}} - \left( E_{iz} - E^{*}_{iz} \right) R^{*}_{iz} n_e n^{*} \\
    & -E_{n,\ce{rad}} R_{n,\ce{rad}} n_e n_{\ce{He}} - E^{*}_{n,\ce{rad}} R^{*}_{n,\ce{rad}} n_e n^{*} \\
    & -E^{*}_{iz} X_{\ce{(GS)\rightarrow (MS)}} n_e n_{\ce{He}} - E_{i,\ce{rad}} R_{i,\ce{rad}} n_e n_{\ce{He+}} - \Gamma_{e,\ce{wall}} T_e n_e \\
    & - \frac{3}{2} \frac{2m_e}{m_i} R_{i,\ce{el}} n_{\ce{He+}} n_e \left(T_e - T_i \right) \\
    & - \frac{3}{2} \frac{2m_e}{m_n} R_{n,\ce{el}} \left(n_{\ce{He}} + n^{*}  \right) n_e \left(T_e - T_n \right) - \frac{3}{2} \frac{dn_{\ce{He+}}}{dt} T_e
\end{split} 
\label{E4}
\end{equation}

The meaning of the coefficients which appear in the balance equations above is summarized in table \ref{tab:0D_model_symbols}. Note that it is assumed that puffing and plasma recycling only produce GS He atoms, while turbo-pumps act as sinks also for the MS population. 

The input parameters of the 0D model are the external power $P_{ext}$, the gas puff strength and pumping speed, which determine $\Gamma_{n,\ce{puff}}$ and $\Gamma_{n,\ce{pump}}$ according to the expressions valid for the unresolved point model contained in \cite{tonello_point_2021}. Moreover, since the expression for the ion-to-wall sink is retrieved from the integration of SOLPS-ITER equations, it is required to specify a radial decay length for the density $\lambda_n$ - mimicking the decay boundary condition scheme - and a cross-field diffusion coefficient $D_{\perp}$ for its computation according to the formula derived in reference \cite{tonello_point_2021}.

\begin{table}[]
\begin{tabular}{lll}
\hline
Symbol                          & Physical meaning                                            & Units                \\ \hline
$R_{iz}$                        & Ionization rate: \ce{e- + He -> 2e- + He+}                  & (cm$^{-3}$ s$^{-1}$) \\
$R^{*}_{iz}$                    & Ionization rate: \ce{e- + He^{*} -> 2e- + He+}              & (cm$^{-3}$ s$^{-1}$) \\
$R_{rc}$                        & Recombination rate: \ce{e- + He+ -> He}                     & (cm$^{-3}$ s$^{-1}$) \\
$R^{*}_{rc}$                    & Recombination rate: \ce{e- + He+ -> He^{*}}                 & (cm$^{-3}$ s$^{-1}$) \\
$R_{n,\ce{rad}}$                & Excitation rate for GS                                      & (cm$^{-3}$ s$^{-1}$) \\
$R^{*}_{n,\ce{rad}}$            & Excitation rate for MS                                      & (cm$^{-3}$ s$^{-1}$) \\
$R_{i,\ce{rad}}$                & Excitation rate for \ce{He+}                                & (cm$^{-3}$ s$^{-1}$) \\
$R_{i,\ce{el}}$                 & Ion - electron elastic collision rate                       & (cm$^{-3}$ s$^{-1}$) \\
$R_{n,\ce{el}}$                 & Electron - neutral elastic collision rate                   & (cm$^{-3}$ s$^{-1}$) \\
$X_{\ce{(GS)\rightarrow (MS)}}$ & Metastable cross-coupling rate: \ce{e- + He -> e- + He^{*}} & (cm$^{-3}$ s$^{-1}$) \\
$X_{\ce{(MS)\rightarrow (GS)}}$ & Metastable cross-coupling rate: \ce{e- + He^{*} -> e- + He} & (cm$^{-3}$ s$^{-1}$) \\
$\Gamma_{i,\ce{wall}}$          & Ion-to-wall sink                                            & (s$^{-1}$)           \\
$\Gamma_{e,\ce{wall}}$          & Electron-to-wall sink                                       & (s$^{-1}$)           \\
$\Gamma_{n,\ce{recyc}}$         & Recycling as GS neutrals at the walls                       & (s$^{-1}$)           \\
$\Gamma_{n,\ce{puff}}$          & Neutral GS gas puff                                         & (s$^{-1}$)           \\
$\Gamma_{n,\ce{pump}}$          & Pumping (both GS and MS)                                    & (s$^{-1}$)           \\
$E_{iz}$                        & GS ionization energy: 24.5 eV                               & (eV)                 \\
$E^{*}_{iz}$                    & MS ionization energy (triplet): 19.8 eV                     & (eV)                 \\
$E_{n,\ce{rad}}$                & Excitation energy for GS                                    & (eV)                 \\
$E^{*}_{n,\ce{rad}}$            & Excitation energy for MS                                    & (eV)                 \\
$E_{i,\ce{rad}}$                & Excitation energy for \ce{He+}                              & (eV)                 \\
$m_e$                           & Electron mass                                               & (A.U.)               \\
$m_n = m_i$                     & Neutral He and \ce{He+} mass (taken as equal)               & (A.U.)               \\
$V = V_e$                       & Volume occupied by neutrals and plasma (taken as equal)      & (cm$^3$)             \\
$P_{ext}$                       & External power absorbed by the plasma                       & (eV s$^{-1}$)        \\ \hline
\end{tabular}
\caption{Summary of the different terms contributing to particle and energy balance equations \ref{E1}, \ref{E2}, \ref{E3}, \ref{E4}.}
\label{tab:0D_model_symbols}
\end{table}

\section*{Acknowledgements}
The authors would like to acknowledge Detlev Reiter, who provided an initial H+He MS-resolved Eirene model. G. Alberti and M. Passoni acknowledge funding from Eni SpA for the PhD program. This work has been carried out within the framework of the EUROfusion Consortium (WP-PWIE), partially funded by the European Union via the Euratom Research and Training Programme (Grant Agreement No 101052200 — EUROfusion). The Swiss contribution to this work has been funded by the Swiss State Secretariat for Education, Research and Innovation (SERI). Views and opinions expressed are however those of the author(s) only and do not necessarily reflect those of the European Union, the European Commission, SERI or ITER Organization. Neither the European Union nor the European Commission nor SERI can be held responsible for them.

\newpage

\section*{Reference}
\bibliographystyle{iopart-num}
\bibliography{Bibliography.bib}  

\end{document}